# Aberration-Free Optical Spectrometer

**Authors:** Qingze Guan[1], Zi Heng Lim[1], Xinchen Wan[1], Yixiu Shen[1], Guangya Zhou[1]*

**Affiliations:**

[1]Department of Mechanical Engineering, National University of Singapore, Singapore,117575

*Corresponding author. Email: mpezgy@nus.edu.sg

**Abstract:** Optical spectrometers are fundamental to scientific analysis, yet achieving high performance at low cost remains challenging because uncorrected aberrations rapidly degrade spectral resolution and typically necessitate complex, expensive optics. Moreover, to preserve spectral resolution, many compact designs remain fundamentally throughput-limited in terms of having a high f-number and a narrow slit. Here we present SHADES (Stochastic High-throughput Aberration-free Deep-Encoded Spectrometer), a general framework that mitigates the effects of optical aberrations using a stochastic grating array (SGA) coupled with physically grounded deep learning (DL), while substantially increasing optical throughput using encoded multi-slits. We develop a theoretical framework establishing aberration resilient spectroscopy in compact, highly-aberrated systems, enabling miniaturization without sacrificing spectral resolution and optical throughput. SHADES utilizes an arbitrary spectrum generator (ASG) for hardware-in-the-loop calibration with a DL-based reconstruction pipeline. We further leverage transfer learning (TL) to reduce calibration data and computation for scalable deployment of SHADES. Experimentally, a μSHADES prototype achieves a spectral resolution of 2.4 nm over 450–700 nm and accurately reconstructs fluorescence spectra for chemical identification. Collectively, SHADES provides an aberration-free, high-throughput, low-cost spectrometer architecture suited for compact and scalable sensing applications.

## Introduction

Grating spectrometers underpin a vast range of scientific and technological applications[1–4], from chemical sensing[5,6] and biomedical imaging[7,8] to astronomy[9,10] and integrated photonics [11,12]. Despite their ubiquity, their performance is fundamentally constrained by optical aberrations. Classical spectrometer designs must carefully balance spectral resolution, optical throughput, size, and tolerance to fabrication and alignment errors, because optical aberrations introduced by imperfections in optics and construction of the spectrometer directly degrade the point spread function (PSF) and spectral fidelity[13–15]. Considerable effort has therefore been devoted to aberration correction through complex optical layouts, precision fabrication, and tight mechanical tolerances, often at the expense of equipment cost, compactness, robustness, and efficiency[8,16,17].

These constraints reflect a deeply rooted assumption: that aberration-free, diffraction-limited performance in a grating spectrometer requires precise control of the optical wavefront[14,17–19]. As a result, aberrations, whether arising from design compromises, miniaturization, fabrication imperfections, or misalignment, are treated as fundamentally detrimental and must be minimized optically[14,20,21]. This assumption enforces long-standing trade-offs between spectral resolution and optical throughput and severely limits the scalability and robustness of compact spectrometers[21–23].

Here, we introduce SHADES (Stochastic High-throughput Aberration-free Deep-Encoded Spectrometer), a fundamentally different paradigm in spectroscopy that decouples spectral resolution from geometric focusing quality. Instead of correcting aberrations, SHADES deliberately randomizes the wavefront using a stochastic grating array (SGA), converting the dispersed optical field into a fully developed speckle pattern that acts as an information-rich encoding of the input spectrum. The resulting diffraction patterns, although highly distorted, retain complete undegraded spectral information. We develop a theoretical framework showing that SHADES delivers diffraction-limited, aberration-free spectral performance independent of the magnitude or type of optical aberrations present in the system. Hence, SHADES enables high-performance optical spectrometers intrinsically immune to classical optical aberrations, fabrication errors, and alignment tolerances. Additionally, the randomized diffraction from the SGA further enables a coded multi-slit entrance aperture based on a binary pseudo-random sequence, further providing SHADES with high optical throughput without sacrificing spectral resolution.

Reconstruction of spectra from SHADES is performed using a deep learning (DL) framework trained on the datasets generated by an in-house developed arbitrary spectrum generator (ASG). This unique hardware-in-the-loop data engine produces a large amount of high-fidelity spectral training datasets that enable accurate and robust spectra reconstruction. Using SHADES, we further show a device-to-device transfer learning (TL) scheme that reduces calibration overhead, thus supporting the deployment of large-scale, aberration-free, high-performance spectroscopic sensors at extremely low cost. Our results establish a new class of spectrometers in which performance is decoupled from optical perfection, opening a pathway toward compact, low-cost, robust, and high-performance spectroscopic instruments that defy traditional design limits.

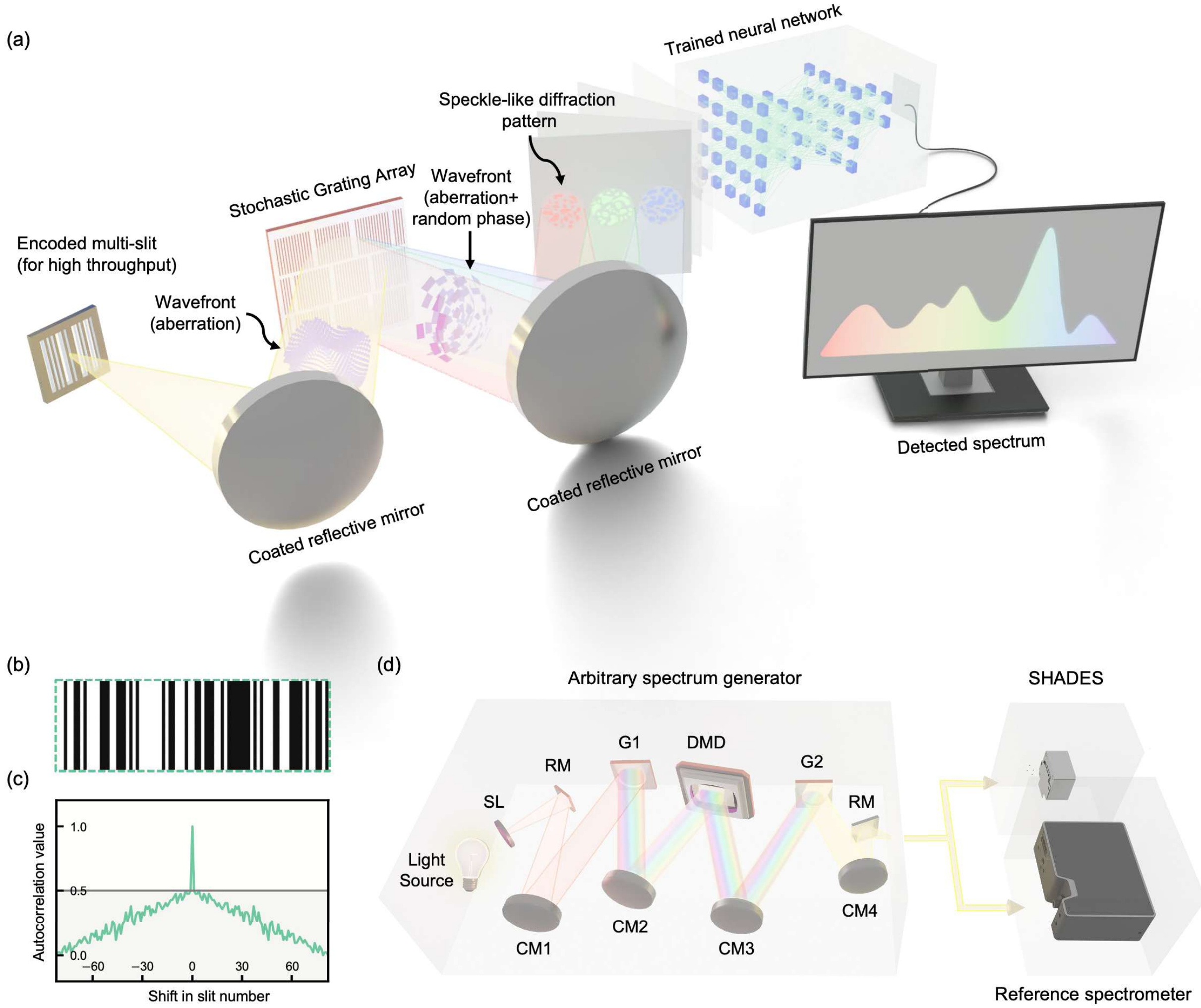

**Fig. 1. SHADES concept and ASG-based calibration. (a)** Working principle of SHADES. An SGA imposes a spatially random phase on the dispersed wavefront, converting low-spatial-frequency aberrated phase distortions into a high-contrast speckle-encoded intensity pattern that carries spectral information. A DL neural network reconstructs the input spectrum from the measured pattern. **(b)** Example of an 83-slit aperture pattern. **(c)** Autocorrelation of the aperture pattern in **(b)**. **(d)** Hardware-in-the-loop calibration workflow. The ASG generates a large library of light signals with diverse spectral profiles, whose outputs are simultaneously measured by both SHADES and a reference spectrometer to obtain paired training data, with the latter serving as the spectral reference. The ASG consists of a slit (SL), reflective mirrors (RM), gratings (G1–G2), a DMD, and concave mirrors (CM1–CM4).

## SHADES framework

SHADES addresses the spectral resolution bottleneck imposed by optical aberrations and limited optical throughput by transforming the spectrometer from a direct-imaging system into a stochastic information encoder/decoder (Fig. 1a). In our reflective spectrometer design, we replace the conventional grating with an SGA, which is formed by partitioning a grating into micro grating cells, each imparting an independent stochastic phase shift by randomly offsetting the micro grating along a direction perpendicular to its grating lines. We show that the spatially stochastic phase shifts on the SGA are transferred to the dispersed wavefronts, and this modulation possesses a unique property that is invariant with respect to both wavelength and field-of-view (Supplementary Note 1) [24]. Hence, the slowly spatially varying aberrated wavefronts are scrambled and converted to high-spatial-frequency, high-contrast and randomized wavefronts. As a result, the PSF becomes a random speckle pattern. The size of the overall PSF intensity envelope is primarily determined by the micro grating cell size, whereas the speckle grain statistics are governed by diffraction of the imparted stochastic phase shifts (Supplementary Note 1).

Our theoretical analysis (Supplementary Note 2) quantifies the spectral resolution of SHADES using the average speckle size ($SS_{avg}$) of the PSF, which estimates the finest resolvable structure in the encoded spectral measurement. We define the $SS_{avg}$ by the full width at half maximum (FWHM) of the normalized autocovariance function (NAF) of the speckle intensity along the dispersion direction. Importantly, our analysis shows that the $SS_{avg}$ is diffraction-limited by the system's aperture and scales approximately as $SS_{avg} \sim \lambda f/a$, where $a$ is the effective aperture size of SHADES, $\lambda$ is the wavelength, and $f$ is the imaging focal length. While optical aberrations can broaden the overall intensity envelope of the PSF, they do not substantially change the $SS_{avg}$. In other words, the spectral resolution of SHADES remains diffraction-limited and is not affected by the system's optical aberrations.

Besides aberration-free spectroscopy, SHADES also enables high optical throughput via coded aperture multiplexing. In conventional dispersive spectrometers, widening the entrance slit increases photon flux but effectively convolves the spectrum with a broader aperture kernel, thus degrading spectral resolution. In SHADES, we employ a spatially distributed multi-slit array patterned according to a pseudo-random binary sequence, such as a maximum length sequence (m-sequence)[25]. The impulse response of SHADES from a single wavelength thus becomes a superposition of PSFs from multiple slit positions. Accordingly, we quantify the spectral resolution of SHADES using the $SS_{avg}$ of the aggregated PSFs (Supplementary Note 3).

A key consideration here is the correlation between encoded PSFs at different slit positions on the entrance aperture. In the ideal limit where the speckle patterns from different slits are mutually uncorrelated, meaning lateral translation produces an independent speckle pattern, then the $SS_{avg}$ and thus the spectral resolution of SHADES remains unchanged, even with its throughput significantly enhanced by the multi-slit entrance (Supplementary Note 3). Conversely, if the PSFs are fully correlated, meaning the PSF is identical at different slit locations, the $SS_{avg}$ is broadened by the multi-slit aperture; this is the worst case for spectral resolution (Supplementary Note 4). In practice, although the stochastic phase profile imposed by the SGA is fixed and independent of wavelength and field angle, field-dependent optical aberrations can progressively decorrelate the PSFs, yielding partially correlated behaviors (Supplementary Note 5). While this decorrelation may benefit spectral resolution, here we bound the system's performance by evaluating it under the conservative limit of full spatial correlation, where the PSFs are assumed to be identical.

Under this assumption, the total NAF of the multi-slit SHADES reduces to the convolution of the single point (pinhole) NAF and the autocorrelation function of the entrance aperture (Supplementary Note 4). For a wide slit, this convolution broadens the response and degrades spectral resolution[26]. In SHADES, by selecting a pseudo-random binary code for the multi-slit entrance, the autocorrelation of the aperture preserves a narrow peak comparable to that of a single slit within the coded slit array. As a result, the $SS_{avg}$, and hence the spectral resolution derived from the total NAF is only determined by the smallest slit width. Fig. 1b illustrates a mult-slit entrance aperture and Fig. 1c shows the autocorrelation of it. This enables a huge increase in optical throughput while maintaining the spectral resolution set by the diffraction-limited speckle grain size. Consequently, SHADES operates in a regime where high optical throughput and high spectral resolution can be achieved simultaneously.

Because the recorded speckle pattern encodes the input light spectrum through a complex forward model, we reconstruct the spectra using DL. However, the exact system response depends on fabrication and alignment tolerances that are difficult to model accurately, making purely simulated training data insufficient. We therefore construct an ASG to generate light with high-quality spectral profiles in large quantities (Fig. 1d). For each synthesized light output from ASG, the SHADES measurement is recorded, while a reference spectrometer simultaneously acquires the ground-truth spectrum, forming paired data for automated DL training with no manual intervention.

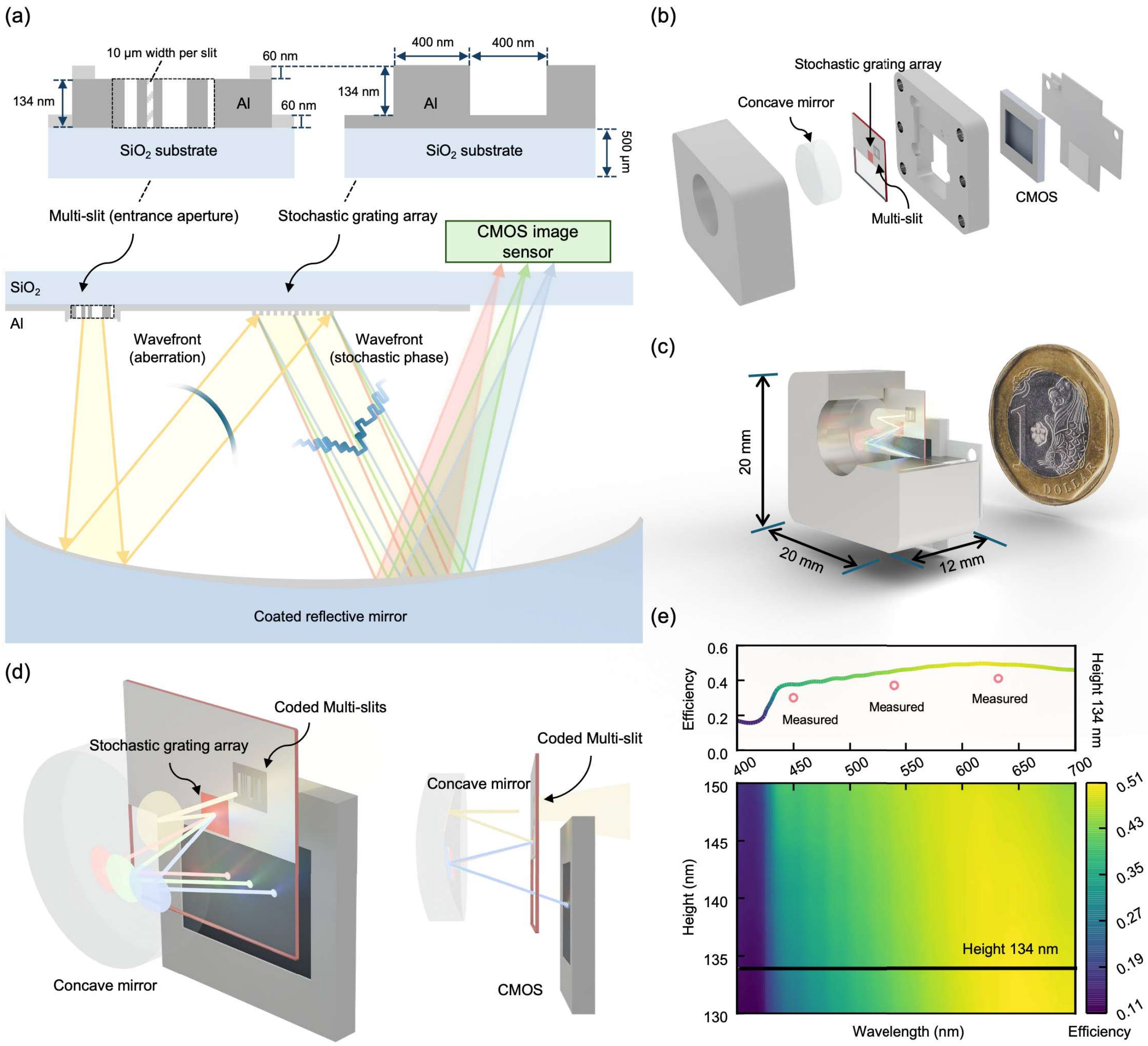

**Fig. 2. Design of the μSHADES prototype. (a)** Cross-sectional schematic of the μSHADES optical architecture. The coded multi-slit entrance aperture and SGA are monolithically integrated on a single fused-silica ($SiO_2$) chip using patterned Al layers in a compact folded reflective geometry. The entrance aperture of μSHADES contains an array of encoded slits, each with a slit

width of 10 μm. Each grating element in the SGA has a period of 800 nm and a groove depth of 134 nm. A uniform 60 nm Al layer is deposited to block unwanted light from entering μSHADES. **(b)** Exploded view of μSHADES components, comprising the enclosure, concave mirror, monolithic chip, and CMOS sensor. **(c)** Assembled μSHADES prototype next to a Singapore one-dollar coin, illustrating its compact form factor. **(d)** Schematic of the off-axis optical design of μSHADES. **(e)** Diffraction efficiency optimization. Heatmap: FDTD-simulated efficiency versus grating and wavelength; top: comparison between simulation and experiment at the selected grating groove depth.

**Design, fabrication and characterization of μSHADES**

Based on our SHADES theory, we design μSHADES, a high-performance miniature spectrometer. Because SHADES is free from optical aberrations, the optical layout can be significantly simplified and does not require multi-element corrected optics and tight fabrication and alignment tolerances. Thus, μSHADES can be compact and extremely low-cost, yet maintain diffraction-limited spectral resolution and high optical throughput. As shown in Fig. 2a and 2d, μSHADES adopts a compact, off-axis folded reflective optical design (detailed design in Supplementary Note 6). Light that enters through the coded multi-slit aperture is collimated by a coated concave mirror, dispersed and phase-scrambled by the SGA, and then reimaged onto a CMOS sensor. We adopt a monolithic fabrication strategy in μSHADES: both the multi-slit entrance aperture and the SGA are lithographically defined on a single fused-silica ($SiO_2$) chip using shared lift-off deposited aluminum (Al) layers (Supplementary Note 7). The resulting μSHADES comprises only three functional optical elements: the micro-fabricated chip, a concave mirror, and a commercial CMOS sensor, packaged in a simple enclosure measuring only $20 \times 20 \times 12$ mm$^3$ (Fig. 2b). The overall device is smaller than a coin (Fig. 2c).

The SGA consists of $40 \times 40$ micro gratings, each with a size of $25 \times 25$ μm$^2$ and shifted laterally along a direction perpendicular to its grooves to introduce a random phase. The micro gratings have a period of 800 nm and a duty cycle of 50%. To maximize the diffraction efficiency of the micro gratings in the SGA, we optimize the grating groove depth using finite-difference time-domain (FDTD) simulations. Fig. 2e is a heat map showing diffraction efficiency as a function of groove depth and wavelength. We focus on p-polarized rather than s-polarized light for high and even efficiency in the functional wavelength range (Supplementary Note 7). We select a groove depth of 134 nm for its relatively high diffraction efficiency throughout the operational wavelength band. Experimental measurements using three lasers, represented by circular symbols in Fig. 2e, show good agreement with the FDTD simulations.

Optical aberrations remain uncorrected, as correction is unnecessary for this application. μSHADES is able to recover high-resolution spectra from this highly aberrated optical setup. As shown on the left in Fig. 3a, the raytracing results show that the wavefront exhibits strong aberrations in our off-axis miniature spectrometer design with a normal grating of the same grating period. Replacing the normal grating with the SGA imposes a random phase profile and scrambles the aberrated wavefront as shown in the right column in Fig. 3a. In the far field, this reshapes the aberrated PSF into a speckle-like one as shown in Fig. 3e. We further quantify this transformation by comparing the modulation transfer function (MTF) of the miniature spectrometer with a normal grating and with an SGA. With a normal grating, the PSF is smeared by optical aberrations, causing the MTF to decay rapidly and greatly suppresses high spatial frequency information. In contrast, μSHADES with an SGA yields an MTF that remains substantially broader, retaining significant values in the high spatial frequency regime. This enhancement is further summarized by the comparison of log-scale MTF cross-section plots along its center (Fig. 3j). Further, the ratio of the SGA-to-normal-grating MTF values (Fig. 3k) and its zoomed-in view (Fig. 3l) reveal a marked enhancement at high spatial frequency, where stochastic encoding using the SGA outperforms the normal grating baseline by multiple times for most frequencies.

Experimental PSF measurements confirm these simulations (Fig. 3e, right). We fabricated and tested both the normal grating and the SGA in the same off-axis miniature spectrometer. The measured PSF at 540 nm wavelength for the normal grating appears as a featureless blur, whereas the SGA yields a high-contrast speckle pattern with fine spatial structure. These results support our theoretical framework that imposing a random phase profile on the aberrated wavefront with an SGA preserves high spatial frequency content in SHADES and improves computational spectrum reconstruction.

The microfabricated fused-silica chip is shown in Fig. 3b. The entrance aperture of μSHADES consists of an array of slits patterned according to a pseudo-random binary sequence (Fig. 3c, d), where 8 out of 15 slits are open and the remaining slits are closed. Each slit has a dimension of 10 μm (width) × 500 μm (height). This coded multi-slit design increases the optical throughput by approximately 400 times compared with a 10 μm × 10 μm square pinhole, while preserving a narrow correlation width required for spectral discrimination. The total NAF of the μSHADES is calculated and plotted (Fig. 3g) as a function of the lateral displacement $\Delta x$ along the dispersion direction on the image plane (Supplementary Notes 5 & 6). The total NAF exhibits a sharp dominant central peak at $\Delta x = 0$ with decaying sidelobes. According to our theory, the spectral resolution of SHADES is established by the FWHM of the total NAF, following its conversion into wavelength units based on the system dispersion. For the present μSHADES geometry, the optical design uses a 4.5 mm focal-length concave mirror with an effective f-number of approximately 3.0, giving a linear spectral dispersion of 1.6 mm on the image sensor over the 450–700 nm wavelength range. As shown in Fig. 3g, the main lobe of the total NAF value drops rapidly to 0.55 at a lateral displacement of approximately 10 μm. This indicates that an estimated spectral resolution of our current μSHADES is 1.55 to 3.1 nm (Supplementary Note 4). By contrast, for a conventional miniature spectrometer with the same optical design and a normal grating, the aberrated spot size near the central wavelength is on the order of 120 μm. This corresponds to an effective spectral resolution of 19 nm—a figure that often degrades further in practice due to fabrication and alignment errors. These results demonstrate that the multi-slit coded entrance aperture is designed not only to maximize optical throughput but also to preserve a sufficiently narrow FWHM of the total NAF. This ensures that adjacent spectral channels remain distinguishable following computational reconstruction. SEM images of the SGA (Fig. 3f) confirm the intended fabrication of these micro gratings and the imposed stochastic initial offsets. The manufactured μSHADES device and its compartments are shown in Fig. 3h and Fig. 3i.

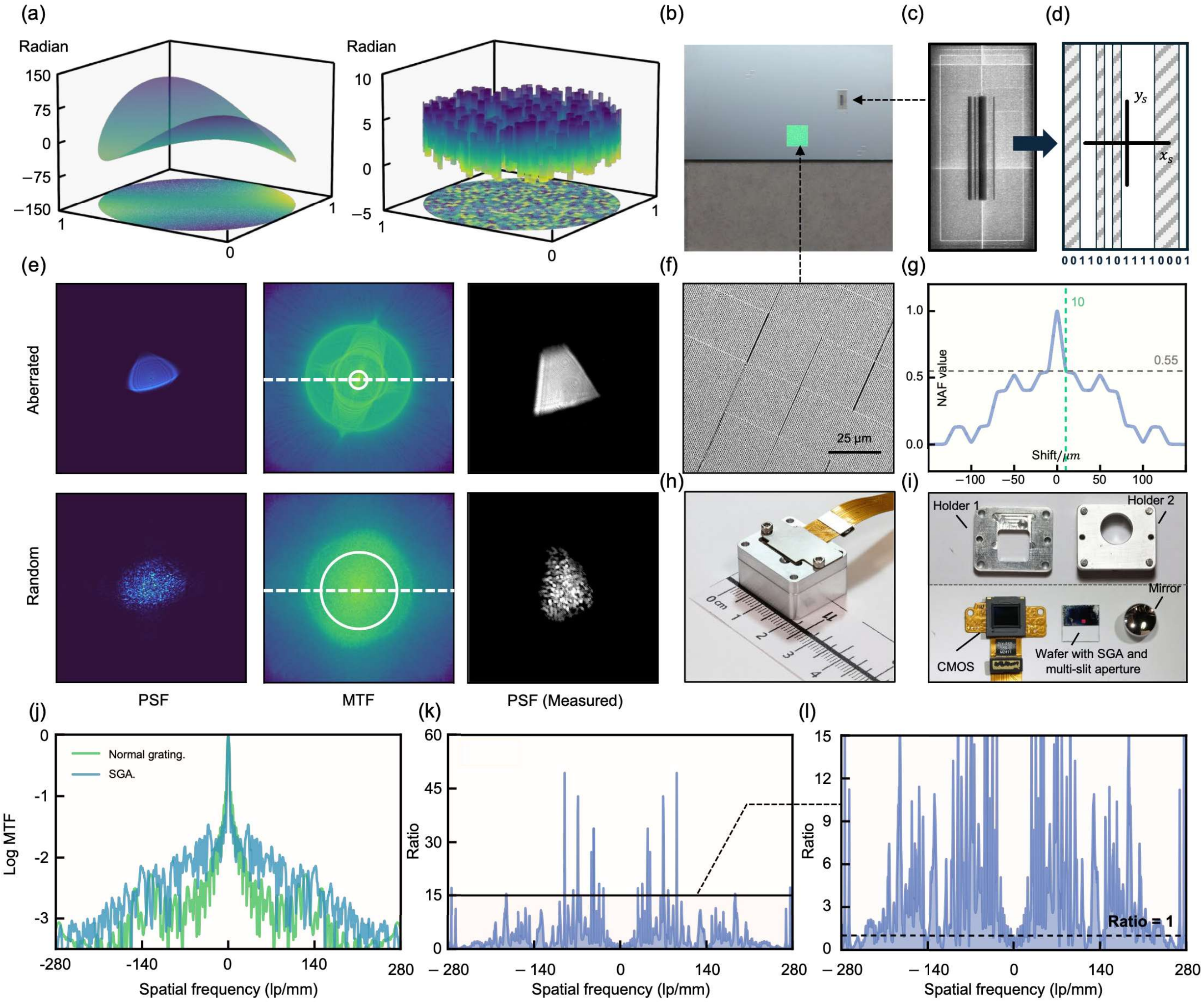

**Fig. 3. Fabrication and characterization of the µSHADES prototype. (a)** Wavefront aberration in the off-axis miniature spectrometer with a normal grating (left) and the randomized phase profile with an SGA (right). **(b)** Photograph of the microfabricated fused-silica chip, where the coded multi-slit aperture and SGA are defined on the upper half of the chip, whereas the lower half is the transparent $SiO_2$ substrate. **(c)** SEM image of the fabricated multi-slit aperture. **(d)** Schematic of the pseudo-random binary sequence used to code the multi-slit aperture. As shown in the figure, "1" denotes an open slit, while "0" denotes a closed slit. **(e)** Optical performance comparison at 540 nm wavelength for the miniature spectrometer with a normal grating (top row) and an SGA (bottom row). Columns display the simulated PSF, simulated MTF, and experimentally measured PSF. The white circles in the MTF plots indicate the spatial frequency boundary where the contrast drops to 1% of the zero-frequency DC component. **(f)** SEM image of the fabricated SGA. **(g)** Total NAF of µSHADES with multi-slit aperture encoded with a pseudo-random binary sequence of length 15. **(h)** Photograph of the assembled µSHADES. **(i)** Core components of µSHADES. **(j–l)** Quantitative MTF analysis: **(j)** log-scale MTF cross-section plots along the center (as indicated by the dashed line in **(e)**) for a normal grating and an SGA, **(k)** the MTF ratio of the SGA to the normal grating configuration, and **(l)** a zoomed-in view of (k), demonstrating the significant recovery of high-spatial-frequency information by using an SGA in µSHADES.

## ASG-based calibration and deep-learning-based reconstruction

We use a deep neural network (DNN) to decode the complex speckle patterns measured from μSHADES and reconstruct spectra. We further develop a hardware-in-the-loop calibration engine that generates a large quantity of high-fidelity paired datasets for DNN training. The core of this engine is the ASG shown in Fig. 4a, which employs a dispersion-encoding-recombination optical layout[27]. A broadband light source passing through a slit is dispersed by a grating and imaged onto a digital micromirror device (DMD). The columns of micromirrors on the DMD map to spectral wavelengths, while the number of "ON" state micromirrors in each column maps to intensity (Supplementary Note 8; Video 1). By dynamically modulating the "ON"/"OFF" states of micromirrors, the ASG synthesizes and outputs light with arbitrary spectrum profiles (Video 2). To correct the intensity nonuniformity of the dispersed slit images on the DMD and to achieve precise control over wavelength and intensity, we use a column-wise single pixel imaging-based[28] calibration method (Supplementary Note 9) and subsequently apply an optimization-based method to generate the DMD pattern for the desired spectrum output (Supplementary Note 10). The synthesized light output from ASG is then split into two paths: one illuminates μSHADES to capture the measurement, and the other is measured by a reference spectrometer to provide the ground-truth spectrum (Supplementary Note 11).

For spectral reconstruction, we implement a ResNet-style[29] model (Fig. 4b). The architecture consists of a convolutional encoder followed by a fully connected decoder. The encoder comprises a series of convolution-batch normalization-LeakyReLU (CBL) blocks organized into residual units. The 2D measurement from μSHADES is first transformed into a high-dimensional feature representation by the residual CBL blocks and then flattened. The flattened features are processed by a three-layer fully connected decoder that regresses the 1D spectrum vector.

We evaluate the trained μSHADES on held-out validation data generated from the ASG, spanning both sparse (narrowband) and dense (broadband) signals. The reconstruction of random mixtures of narrow spectral lines is illustrated in Fig. 4c. The system accurately retrieves peak positions and relative intensities, achieving a measured FWHM spectral resolution of 2.4 nm in the 450–700 nm range. Furthermore, the system exhibits high fidelity in reconstructing continuous, broadband spectra with complex features (Fig. 4d, additional results in Supplementary Note 12). Notably, the current resolution might be limited by the output spectral resolution of the ASG rather than the intrinsic encoding limit of the μSHADES. Finer resolution may be unlocked with a higher resolution ASG. To isolate the contribution of stochastic-phase encoding using an SGA, we built and calibrated a control spectrometer identical in optical design and training protocol but using a normal grating with the same grating period instead of an SGA. Under the same dataset and network settings, the control system failed to recover spectra with comparable fidelity, underscoring the necessity of the SGA (Supplementary Note 13).

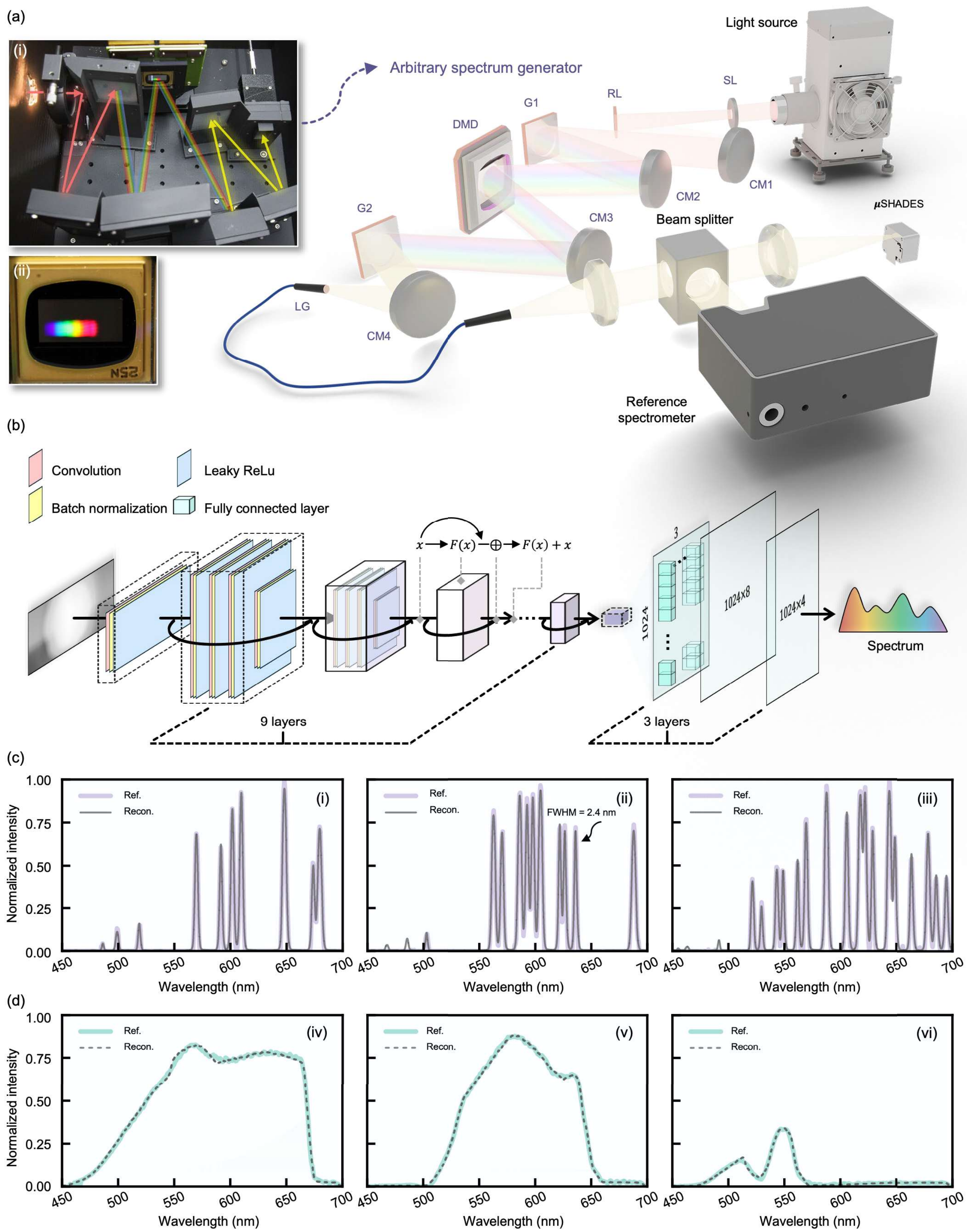

**Fig. 4. Hardware-in-the-loop calibration and spectral reconstruction performance. (a)** Schematic of the ASG. The system employs a dispersion-encoding-recombination optical layout for precise spectral synthesis. Light from an incoherent source passes through an entrance slit (SL),

is reflected by a reflective mirror (RM), collimated by a concave mirror (CM1) onto a diffraction grating (G1), and the dispersed spectrum is imaged by CM2 onto a DMD for wavelength-selective modulation. The modulated light is collected by CM3, recombined by diffraction grating G2, and coupled into an output light guide (LG) via CM4. The output light is split to illuminate both μSHADES and the reference spectrometer. Insets: (i) photograph of the ASG and (ii) dispersed slit images projected on the DMD. **(b)** Architecture of the DNN. The model utilizes a residual learning framework comprising an encoder with nine layers of CBL blocks and skip connections. **(c)** Reconstruction results for sparse spectra. Comparison between the ground truth and μSHADES reconstruction for random narrowband peaks demonstrates μSHADES' ability in the accurate retrieval of peak positions and intensities. **(d)** Reconstruction results for broadband spectra. Results again demonstrate μSHADES' ability in accurate recovery of complex continuous spectra, validating the network's generalization capability across different spectral profiles.

**Evaluation of the ASG-trained μSHADES and transfer learning**

To validate the operational robustness of the ASG-trained μSHADES in real application scenarios, we test the system using various independent physical light sources and filters. As shown in Fig. 5a, μSHADES successfully reconstructs the discrete emission lines of a spectral calibration lamp with a FWHM of 2.4 nm, matching the spectral resolution limit established during ASG training. Testing results with narrowband filters and multi-color LEDs in Fig. 5b and 5c further confirm the system's robustness and accuracy in recovering both sharp spectral features and continuous broadband spectra, indicating that a SHADES model trained solely on ASG-generated data generalizes effectively to practical, diverse spectra.

Our ASG calibration and training framework uses incoherent illumination, where the intensities of light sum over the detector rather than amplitudes[30]. This configuration is in line with most spectroscopic sensing applications[31–35]. However, problems may arise when it is used to sense highly coherent sources like lasers. These light sources have stable phase relationships and the amplitudes of light sum over the detector, producing laser speckle patterns. This creates a domain shift between the coherent test dataset and the incoherent training dataset, leading to reconstruction failure. We tested the ASG-trained μSHADES with a 540 nm laser, as shown in Fig. 5d, and the PSF recorded together with its corresponding failed reconstruction result is highlighted in the figure. To bridge this gap, we introduced a rotating diffuser into the optical path between the source and μSHADES to scramble the spatial coherence (Supplementary Note 14). This effectively averages the varying coherent speckle over the integration time of the image sensor, synthesizing an incoherent response that aligns with our calibration/training model. Under these conditions, the laser spectrum is accurately recovered as shown in Fig. 5d, demonstrating that μSHADES can handle coherent sources via simple decoherence techniques. We emphasize that the coherence of light needs to be considered in the design of computational spectrometers. Calibration with an incoherent light source and the usage of a rotating diffuser collectively enable μSHADES to function under both coherent and incoherent light conditions.

For scalable manufacturing and mass deployment of μSHADES as spectroscopic sensors, per-device calibration is a major cost[36,37]. Fabrication tolerances can introduce minor variations in μSHADES units, including differences in grating position $z$ and rotation $\theta$ (Fig. 5e, inset), which modify the speckle response across the individual units. We therefore implement a TL strategy. Rather than collecting a full dataset and training for every device, we leverage the weights learned from one device and fine-tune them on a target device using a fraction of the data. As shown in Fig. 5e, TL reaches near-baseline performance using 10% of the target dataset, whereas training from scratch on the same reduced dataset converges poorly. This approach reduces calibration effort by ~90% (Supplementary Note 15).

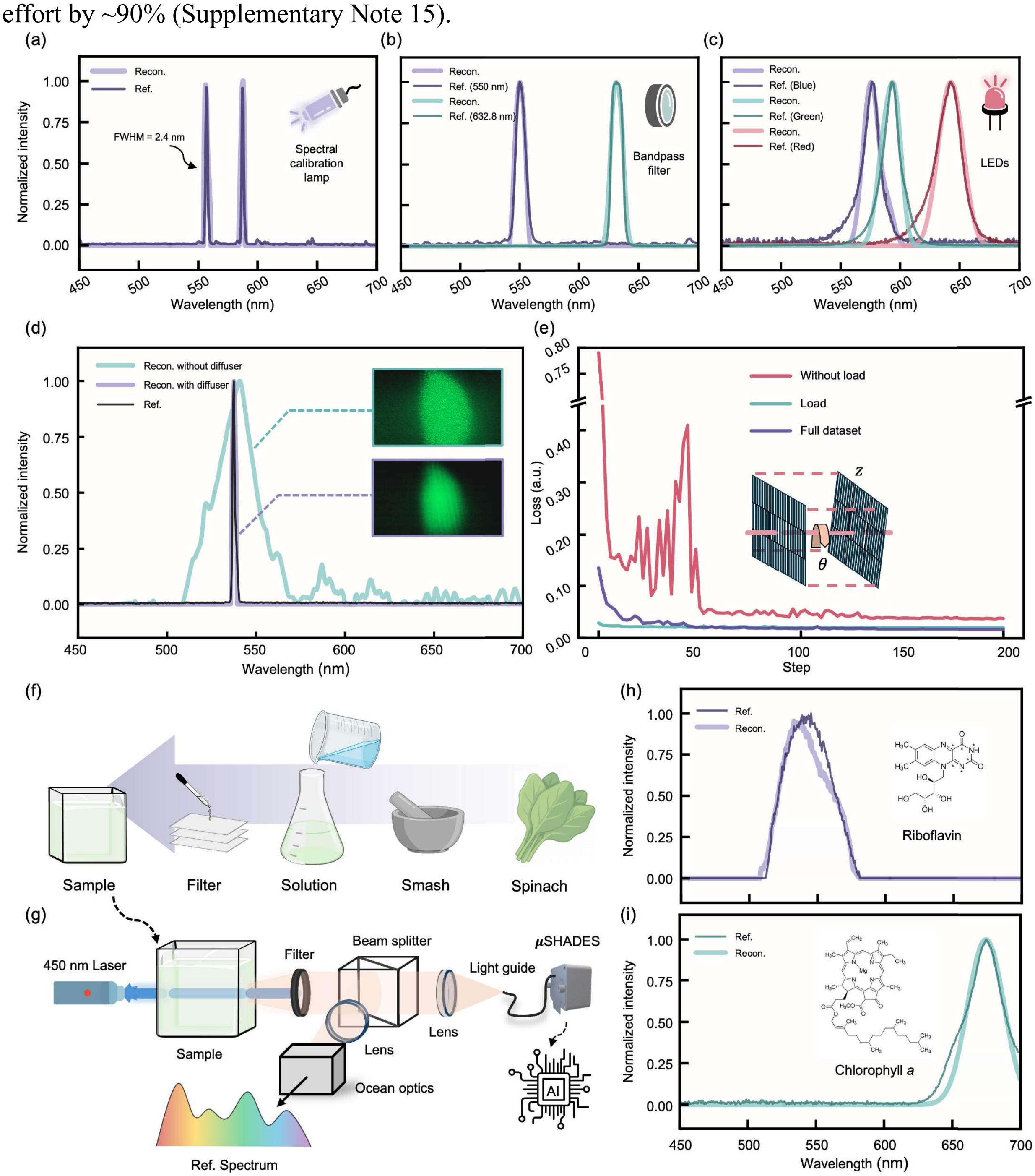

**Fig. 5. Experimental validation of μSHADES and scalable deployment of spectroscopic and fluorescence sensing. (a–c)** Spectral reconstruction of diverse independent light sources. Comparisons between μSHADES reconstructions and reference spectra for **(a)** a spectral calibration lamp exhibiting discrete lines, **(b)** bandpass filters, and **(c)** broadband multi-color LEDs. μSHADES demonstrates a spectral resolution of 2.4 nm. **(d)** Coherence effect testing results. Direct illumination of a laser generates a high-contrast static speckle field (top inset) that does not

match the incoherent dataset generated from ASG, resulting in reconstruction failure. Inserting a rotating diffuser temporally averages the varying coherent speckle to synthesize an incoherent response (bottom inset), enabling accurate spectral recovery. **(e)** Scalable training via TL. The inset illustrates typical manufacturing tolerances (grating position $z$, rotation $\theta$) that create unit-specific PSF variations. The plot compares training loss curves: a model initialized via TL converges to the full-dataset baseline using only 10% of the data, whereas training from scratch on the limited dataset fails to converge. **(f, g)** Fluorescence measurement workflow. **(f)** Sample preparation for pigment extraction (smashing, dissolution, and filtration). **(g)** Optical setup using 450 nm excitation and μSHADES detection of fluorescence emission. **(h, i)** Biochemical sensing results. Reconstructed fluorescence spectra of **(h)** riboflavin (vitamin $B_2$) and **(i)** chlorophyll *a* are in agreement with fluorescence spectra obtained with the reference spectrometer.

Finally, we demonstrate the application of μSHADES in portable biochemical sensing (Fig. 5f, g). We prepare solutions of riboflavin (vitamin $B_2$)[38] and chlorophyll *a* [39] from spinach samples. Using a 450 nm excitation laser paired with a long pass filter to reject the laser wavelength, we capture the fluorescence emission spectra of the samples. μSHADES accurately resolves the characteristic emission band of riboflavin centered at 550 nm (Fig. 5h) and the red fluorescence of chlorophyll *a* near 670 nm (Fig. 5i). These results, validated against the spectra obtained by a commercial reference spectrometer, highlight the potential of μSHADES as a low-cost, portable module for in-situ spectroscopic sensing applications, including food safety[40] and environmental monitoring[41].

## Conclusion

We introduce SHADES, a generalized stochastic-encoding framework for spectroscopy that renders spectral resolution largely insensitive to optical aberrations and construction errors, and at the same time, enables high optical throughput using a coded multi-slit aperture. The key component enabling SHADES is the SGA that imparts a random phase profile, independent of incident wavelengths and field angles, to the aberrated wavefronts. A theoretical framework for SHADES is also established, showing its ability to deliver diffraction-limited, aberration-free spectral sensing performance independent of the magnitude or type of optical aberrations present in the system.

We implement SHADES in μSHADES, a miniature spectrometer built around a monolithic chip that integrates the coded multi-slit aperture and SGA. To overcome simulation-to-reality mismatch and deployment-dependent perturbations, we develop a hardware-in-the-loop calibration/training engine based on an ASG that produces a large quantity of high-quality paired datasets. Using the ASG-trained μSHADES, we achieve a spectral resolution of 2.4 nm over 450–700 nm and demonstrate generalization to diverse independent light sources. We further validate biochemical sensing through fluorescence measurements of riboflavin and chlorophyll *a*, showing close agreement with their reference spectra.

Together, these results establish a practical route to miniaturized spectroscopy that is aberration-free, easily manufacturable, high-throughput, and low-cost. The SHADES framework is compatible with diverse optical layouts and can be advanced through higher-resolution spectral synthesis for calibration and task-specific reconstruction models. We anticipate applications in field-deployable chemical analysis, food safety, and biomedical sensing, where size, cost, and photon efficiency are critical.

**Methods**

Optical design and simulation

The μSHADES prototype adopts an off-axis Ebert-type spectrometer geometry. The optical layout is modeled and optimized in Zemax OpticStudio. A coded multi-slit mask is used as the entrance aperture (aperture size 150 μm × 500 μm consisting of 15 slits, 8 open and 7 closed, each of size 10 μm × 500 μm). Light from the aperture is collimated by a concave mirror (f = 4.5 mm) and directed onto a reflective SGA (40 × 40 cells, active area 1 mm × 1 mm). The dispersed wavefront is reflected back to the same concave mirror and imaged onto the detector with unity magnification (effective f/# ≈ 3). The detector is a Sony IMX586 CMOS sensor (8000 × 6000 pixels, pixel pitch 0.8 μm, down-sampled and cropped to be 1560 × 512). Images are recorded and cropped to the region containing the speckle-encoded spectral response for subsequent processing.

ASG and experimental setup

For the ASG, the illumination source is an incoherent quartz–tungsten–halogen lamp (Newport 6334NS, 250W). The source light is focused through a slit and dispersed by a grating (Newport 33010FL01-270R) onto a DMD (Vialux V-9501) to enable wavelength-selective modulation. The wavelength-to-DMD mapping is calibrated using a reference spectrometer (Ocean Optics FLAME-T-VIS-NIR-ES).

For measurements, the ASG output is coupled into a 1.5-mm-core optical fiber to promote mode mixing and improve coupling stability. A beam splitter splits the beam into two paths. Two filters (Thorlabs FELH0450 and Thorlabs FESH0700) define the functional spectral range of 450 to 700 nm. One path is measured by the reference spectrometer to record ground truth spectra. The other path is coupled into μSHADES to record the corresponding speckle image. This simultaneous acquisition minimizes temporal drift and source fluctuations.

For independent-source validation, μSHADES is tested using a spectral calibration lamp (Newport 6031 Kr) and LEDs. Bandpass filter measurements are performed at 550 nm (Thorlabs FBH550-10) and 632.8 nm (Thorlabs FL632.8-10). For coherent-source testing, a laser (CNI Laser MGL-F-540) is coupled using a microscope objective. A rotating diffuser mounted on a motorized rotation stage (Thorlabs ELL14) is placed after the laser to remove spatial coherence during measurements.

Dataset acquisition and DNN training

The training dataset comprises 50,000 paired measurements and ground truth spectra. Target spectra are synthesized as random linear combinations of narrowband peaks and broadband backgrounds to approximate diverse source spectra. To reduce temporal persistence and inter-frame coupling, a dark frame (DMD all “OFF”) is inserted between consecutive spectral projections. Acquisition speed is limited by the CMOS exposure time and sensor readout latency.

The dataset is randomly split into 90% training and 10% validation subsets. Each measurement image is recorded as a single-channel grayscale frame and cropped to 1560 × 512 binned pixels.

The DL model consists of an initial convolutional stem followed by a deep feature extractor with nine residual blocks. We use skip connections between CBL blocks. Features are flattened and mapped to the spectrum using a three-layer fully connected regression head. The model is implemented in PyTorch and trained on a workstation equipped with an NVIDIA GeForce RTX 4090 GPU. The batch size is 240. We use the Adam optimizer with a learning rate of $1\times10^{-3}$ for training.

The loss function $\mathcal{L}(\hat{s}, s)$ is a mean absolute error (L1) and mean squared error (MSE) terms to balance sparsity and smoothness in the recovered spectra:

$$\mathcal{L}(\hat{s}, s) = \lambda_1 \frac{1}{N}\sum_{i=1}^{N}|\hat{s_i} - s_i| + \lambda_2 \frac{1}{N}\sum_{i=1}^{N}(\hat{s_i} - s_i)^2$$

where $\hat{s}$ is the reconstructed spectrum, $s$ is the ground truth spectrum, and $N$ is the number of spectral samples. In our implementation, $\lambda_1 = \lambda_2 = 0.5$. Early stopping is used to limit overfitting. Convergence typically occurs within 100 epochs.

Biological sample preparation and measurement

Fluorescence validation experiments are performed using vitamin $B_2$ and chlorophyll *a*. Vitamin $B_2$ is dissolved in deionized water and filtered to remove undissolved solids. Chlorophyll *a* is extracted from spinach using anhydrous ethanol: spinach tissue is homogenized in anhydrous ethanol, and the extract is filtered to remove particulates. The filtration is allowed to settle, and the clarified supernatant is collected for measurement. Solutions are loaded into quartz cuvettes. Excitation is provided by a 450 nm laser diode (Thorlabs CPS450).

Fluorescence emission is collected from the cuvette and passed through a filter set consisting of a 500 nm longpass filter (Thorlabs FELH0500) and a 700 nm shortpass filter (Thorlabs FESH0700). These filters suppress residual pump light. The filtered emission is then collimated and split for simultaneous measurement by μSHADES and the reference spectrometer.

**Acknowledgments:** This work was supported by the Ministry of Education of Singapore (MOE-T2EP50122-0013). **Author contributions:** Conceptualization and theory development: G. Z. Methodology: Q. G., Z. H. L., X. W., G. Z. Investigation: Q. G., Z. H. L., X. W. Visualization: Y. S., Q. G. Supervision: G. Z. Writing – original draft: Q. G. Writing – review & editing: Z. H. L., G. Z. **Competing interests:** Q. G., Z. H. L., and G. Z. are inventors of a patent application on SHADES. **Data, code, and materials availability:** All data needed to evaluate the conclusions in the paper are present in the paper and/or the Supplementary Materials. Additional data are available from the corresponding author upon reasonable request.

**Supplementary Materials**

Supplementary Notes 1 to 16

Figs. S1 to S14

Tables S1 to S3

Supplementary Materials for

# Aberration-Free Optical Spectrometer

Qingze Guan, Zi Heng Lim, Xinchen Wan, Yixiu Shen and Guangya Zhou

Corresponding author: mpezgy@nus.edu.sg

**The PDF file includes:**

Supplementary Notes 1 to 16
Figs. S1 to S14
Tables S1 to S3

# Supplementary Notes

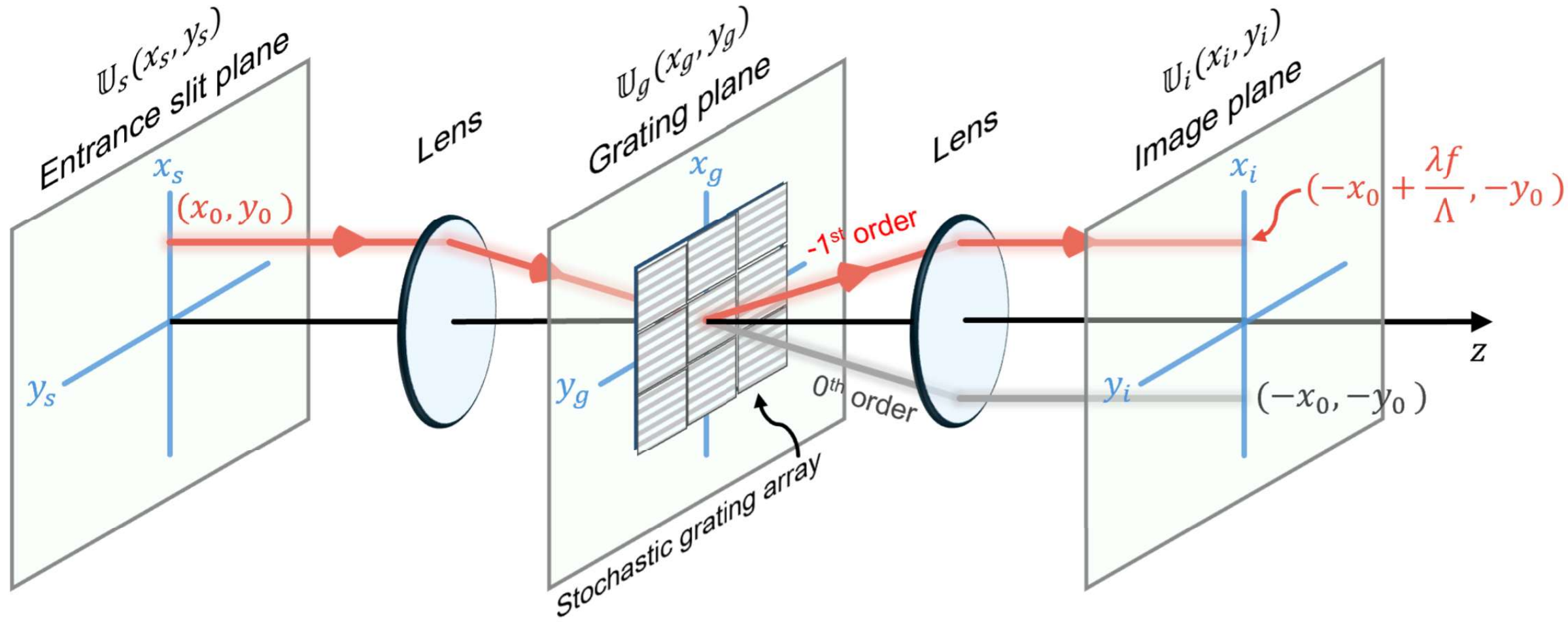

**Fig. S1. Theoretical model of the optical propagation process of typical grating spectrometers.** A point located at $(x_0, y_0)$ on the source plane $U_s(x_s, y_s)$ is relayed to the grating plane $U_g(x_g, y_g)$ and subsequently to the image plane $U_i(x_i, y_i)$. The 0th-order diffraction, shown as the gray ray, is relayed to $(-x_0, -y_0)$. In contrast, the $-1$st-order diffraction, shown as the red ray, is laterally displaced by $\lambda f/\Lambda$ along the $x_i$ axis and forms an image at $(-x_0 + \lambda f/\Lambda, -y_0)$. Here, $\lambda$ is the wavelength, $f$ is the effective focal length of the imaging optics, and $\Lambda$ is the grating period. This schematic illustrates the wavelength-dependent spatial shift that underpins dispersive mapping in a typical grating-based spectrometer.

## Supplementary Note 1. Normal grating spectrometer

As shown in Fig. S1, we consider a $4f$ spectrometer model: the slit plane is located at the front focal plane of a collimator lens (focal length $f$), the grating is placed at its back focal plane (Fourier plane), and the detector is located at the back focal plane of an identical lens. With a common spectrometer model, let $U_s(x_s, y_s)$ be the optical field at point $(x_s, y_s)$ at the slit plane. We used a transmission grating for simplicity of schematic drawing. Throughout the following analysis we assume unit magnification. For the optical field with wavelength $\lambda$, after passing through a lens with a focal length of $f$, the field at the grating plane $U_g(x_g, y_g)$ is:

$$U_g(x_g, y_g) = \frac{e^{ikf}}{i\lambda f} \iint U_s(x_s, y_s) e^{-i\frac{2\pi}{\lambda f}(x_g x_s + y_g y_s)} \, dx_s \, dy_s \tag{S1}$$

where $k = 2\pi/\lambda$ is the wavenumber. The constant term will be ignored in the following analysis.

For a point source at $(x_0, y_0)$ at the entrance slit plane:

$$U_s(x_s, y_s) = \delta(x_s - x_0)\delta(y_s - y_0) \tag{S2}$$

Then the field immediately before the grating becomes

$$U_g^-(x_g, y_g) = e^{-i\frac{2\pi}{\lambda f}(x_g x_0 + y_g y_0)} \tag{S3}$$

We consider the grating has a period of $\Lambda$, grating lines are along $y_g$ direction. Its transmission, hence, is only a function of $x_g$, and periodic in $x_g$, and can thus be expressed in Fourier series. The grating function $g(x_g)$ can be written as:

$$g(x_g) = \sum_{m=-\infty}^{+\infty} a_m e^{i\frac{2\pi m x_g}{\Lambda}} \tag{S4}$$

where

$$a_m = \frac{1}{\Lambda}\int_0^{\Lambda} g(x_g) e^{-i\frac{2\pi m x_g}{\Lambda}} dx_g \tag{S5}$$

We are only interested in $m$-th diffraction order, the field right after the grating is then

$$U_g^+(x_g, y_g) = P(x_g, y_g) U_g^-(x_g, y_g) a_m e^{i\frac{2\pi m x_g}{\Lambda}} \tag{S6}$$

where $P(x_g, y_g)$ is the pupil function at the grating plane.

Suppose the grating is shifted in the $x_g$-axis by a displacement $d$, this is equivalent to adding a phase shift of $2\pi md/\Lambda$ to the $m$-th order diffracted beam, and this can be seen from the Fourier coefficient of the shifted grating:

$$\begin{aligned} a'_m &= \frac{1}{\Lambda}\int_0^{\Lambda} g(x_g - d) e^{-i\frac{2\pi m x_g}{\Lambda}} dx_g \\ &= \frac{1}{\Lambda} e^{-i\frac{2\pi m d}{\Lambda}} \int_0^{\Lambda} g(x_g - d) e^{-i\frac{2\pi m (x_g - d)}{\Lambda}} d(x_g - d) \\ &= e^{-i\frac{2\pi m d}{\Lambda}} \cdot a_m \end{aligned} \tag{S7}$$

It can be seen that the phase shift is independent of wavelength and is also independent of incident angle.

Combining Eq. (S3) and Eq. (S6), the wavefront after the grating is:

$$U_g^+(x_g, y_g) = a_m P(x_g, y_g) e^{-i\frac{2\pi}{\lambda f}\left[\left(x_0 - \frac{m\lambda f}{\Lambda}\right) x_g + y_0 y_g\right]} \tag{S8}$$

After the $2^{nd}$ lens with a focal distance of $f$, field on image plane is then:

$$
\begin{aligned}
U_i(x_i,y_i) &= \iint U_g^{+}(x_g,y_g)e^{-i\frac{2\pi}{\lambda f}(x_ix_g+y_iy_g)}dx_gdy_g \\
&= a_m \iint P(x_g,y_g)e^{-i\frac{2\pi}{\lambda f}\left[\left(x_i+x_0-\frac{m\lambda f}{\Lambda}\right)x_g+(y_i+y_0)y_g\right]}dx_gdy_g \\
&= c\left[\iint P(x_g,y_g)e^{-i\frac{2\pi}{\lambda f}(x_ix_g+y_iy_g)}dx_gdy_g\right]*\delta\left(x_i+x_0-\frac{m\lambda f}{\Lambda},y_i+y_0\right)
\end{aligned}
\tag{S9}
$$

where $c$ is related to $a_m(\lambda f)^2$ is a constant including the $m$-th order diffraction coefficient and the Fourier-transform scaling factor. Hence for incoherent imaging, which is typically true for spectroscopy, the shape of the intensity point spread function (IPSF) is given by the squared magnitude of the coherent amplitude point spread function (APSF):

$$
\mathrm{IPSF}(x_i,y_i) = |\mathrm{APSF}(x_i,y_i)|^2 \tag{S10}
$$

$$
\mathrm{APSF}(x_i,y_i) = \iint P(x_g,y_g)e^{-i\frac{2\pi}{\lambda f}(x_ix_g+y_iy_g)}\,dx_g\,dy_g \tag{S11}
$$

where $\mathrm{APSF}(x_i,y_i)$ is located at $(-x_0+\frac{m\lambda f}{\Lambda},-y_0)$. With $G(x_g,y_g)$ as a rectangular aperture function with size $a$,

$$
G(x_g,y_g) = \mathrm{rect}\left(\frac{x_g}{a}\right)\mathrm{rect}\left(\frac{y_g}{a}\right) \tag{S12}
$$

For aberration-free spectrometers,

$$
P(x_g,y_g) = G(x_g,y_g) \tag{S13}
$$

For aberrated spectrometers,

$$
P(x_g,y_g) = G(x_g,y_g)e^{ikW(x_g,y_g)} \tag{S14}
$$

$W(x_g,y_g)$ is the wavefront aberration of the spectrometer at the grating-plane. On this basis, we start the deduction of our proposed Stochastic High-throughput Aberration-free Deep-Encoded Spectrometer (SHADES).

## Supplementary Note 2. SHADES with single-point input

With the theoretical model for a conventional grating spectrometer established above, we now develop the corresponding theory for SHADES. We first consider a single point source located at $(x_0,y_0)$ on the entrance aperture plane. A stochastic grating array (SGA) is used instead of a single uniform grating. The SGA is a square shape with size of $a$ and is formed by sectioning a normal grating into smaller micro grating with size $b$. Each micro grating is shifted randomly along the $x_g$-axis, thereby adding a stochastic phase shift to the aberrated wavefront. For SHADES, the pupil function now becomes:

$$P(x_g, y_g) = G(x_g, y_g) e^{i[kW(x_g, y_g) + \phi_r(x_g, y_g)]} \tag{S15}$$

$\phi_r(x_g, y_g)$ is the stochastic phase from SGA and is independent of wavelength and ray incident angle. Due to the stochastic phase introduced, the IPSF now becomes a speckle-like intensity pattern.

Let $x_i' = x_i + \Delta x_i$ and $y_i' = y_i + \Delta y_i$, so $(x_i, y_i)$ and $(x_i', y_i')$ stands for two neighboring points. Next, we will estimate the autocorrelation function $\Gamma(x_i, y_i; x_i', y_i')$ for two points $(x_i, y_i)$ and $(x_i', y_i')$ of this speckle pattern:

$$\Gamma(x_i, y_i; x_i', y_i') = \langle \text{IPSF}(x_i, y_i) \cdot \text{IPSF}(x_i', y_i') \rangle \tag{S16}$$

where $\langle \cdot \rangle$ is the ensemble average over SGA. The width of this autocorrelation function provides a reasonable measure of the average speckle size $SS_{\text{avg}}$ of the pattern, which in turn will be used as an estimation of the spectrometer's spectral resolution.

Similar to the laser speckle field, under the fully developed speckle assumption, the field pattern here is a circular complex Gaussian field due to stochastic phase introduced by the SGA. Hence, according to the Siegert relation, the intensity correlation function $\Gamma(x_i, y_i; x_i', y_i')$ for point $(x_i, y_i)$ and $(x_i', y_i')$ is related to the normalized amplitude correlation function through the following equation, assuming $(x_i', y_i')$ is very close to $(x_i, y_i)$:

$$\begin{aligned} \Gamma(x_i, y_i; x_i', y_i') &= \langle \text{IPSF}(x_i, y_i) \rangle \left\langle \text{IPSF}(x_i', y_i') \right\rangle + \left| \left\langle \text{APSF}(x_i, y_i) \text{APSF}^*(x_i', y_i') \right\rangle \right|^2 \\ &= \langle \text{IPSF}(x_i, y_i) \rangle^2 \left[ 1 + \frac{|\langle \text{APSF}(x_i, y_i) \text{APSF}^*(x_i', y_i') \rangle|^2}{\langle \text{IPSF}(x_i, y_i) \rangle^2} \right] \end{aligned} \tag{S17}$$

Let $x_g' = x_g + \Delta x_g$ and $y_g' = y_g + \Delta y_g$ and take Eq. (S11)

$$\begin{aligned} \left\langle \text{APSF}(x_i, y_i) \text{APSF}^*(x_i', y_i') \right\rangle = &\iiiint \left\langle P(x_g, y_g) P^*(x_g', y_g') \right\rangle \\ &\cdot e^{-i\frac{2\pi}{\lambda f}(x_i x_g + y_i y_g)} e^{i\frac{2\pi}{\lambda f}(x_i' x_g' + y_i' y_g')} \, dx_g \, dy_g \, dx_g' \, dy_g' \end{aligned} \tag{S18}$$

$P(x_g, y_g)$ and $P(x_g', y_g')$ follows the definition of Eq. (S14). Due to the SGA introduced, if the points $(x_g, y_g)$ and $(x_g', y_g')$ are not close enough, $P(x_g, y_g)$ is completely uncorrelated to $P(x_g', y_g')$. Hence, the following assumption is adopted:

$$\langle P(x_g, y_g) P^*(x_g', y_g') \rangle = G(x_g, y_g) \mu(\Delta x_g, \Delta y_g) \tag{S19}$$

where $\mu(\Delta x_g, \Delta y_g)$ is a correlation function determined by $b$, which is the size of the micro grating

cells

$$\mu(\Delta x_g, \Delta y_g) = \mathrm{rect}\left(\frac{\Delta x_g}{b}\right)\mathrm{rect}\left(\frac{\Delta y_g}{b}\right) \tag{S20}$$

In fact, if $b$ is infinitely small,

$$\mu(\Delta x_g, \Delta y_g) = \delta(\Delta x_g)\delta(\Delta y_g) \tag{S21}$$

Inserting back into the Eq. (S18), we have

$$\begin{aligned}\left\langle \mathrm{APSF}(x_i, y_i)\,\mathrm{APSF}^*(x_i', y_i')\right\rangle = \iiiint G(x_g, y_g)\,\mu(\Delta x_g, \Delta y_g)\,e^{-i\frac{2\pi}{\lambda f}(x_i x_g + y_i y_g)} \\ \cdot\, e^{i\frac{2\pi}{\lambda f}(x_i' x_g' + y_i' y_g')}\,dx_g\,dy_g\,dx_g'\,dy_g'\end{aligned} \tag{S22}$$

The exponential factor can be simplified as:

$$\begin{aligned}&x_i x_g + y_i y_g - (x_i + \Delta x_i)(x_g + \Delta x_g) - (y_i + \Delta y_i)(y_g + \Delta y_g) \\ &= -x_i\,\Delta x_g - \Delta x_i\,x_g - \Delta x_i\,\Delta x_g - y_i\,\Delta y_g - \Delta y_i\,y_g - \Delta y_i\,\Delta y_g\end{aligned} \tag{S23}$$

For small $\Delta x_g$ and $\Delta y_g$, cross terms like $\Delta x_i \Delta x_g$ and $\Delta y_i \Delta y_g$ are small and can be neglected. Hence, through a change of variables, Eq. (S22) can be rewritten as:

$$\begin{aligned}\langle \mathrm{APSF}(x_i, y_i)\mathrm{APSF}^*(x_i', y_i')\rangle = \iint G(x_g, y_g) e^{-i\frac{2\pi}{\lambda f}(\Delta x_i x_g + \Delta y_i y_g)}\,dx_g\,dy_g \\ \cdot \iint \mu(\Delta x_g, \Delta y_g) e^{-i\frac{2\pi}{\lambda f}(x_i \Delta x_g + y_i \Delta y_g)}\,d\Delta x_g\,d\Delta y_g\end{aligned} \tag{S24}$$

With Eq. (S12),

$$\iint G(x_g, y_g) e^{-i\frac{2\pi}{\lambda f}(\Delta x_i x_g + \Delta y_i y_g)}\,dx_g\,dy_g = a^2\,\mathrm{sinc}\left(\frac{a\Delta x_i}{\lambda f}\right)\mathrm{sinc}\left(\frac{a\Delta y_i}{\lambda f}\right) \tag{S25}$$

And with Eq. (S20),

$$\iint \mu(\Delta x_g, \Delta y_g) e^{-i\frac{2\pi}{\lambda f}(x_i \Delta x_g + y_i \Delta y_g)}\,d\Delta x_g\,d\Delta y_g = b^2\,\mathrm{sinc}\left(\frac{b x_i}{\lambda f}\right)\mathrm{sinc}\left(\frac{b y_i}{\lambda f}\right) \tag{S26}$$

We use the normalized definitions $\mathrm{sinc}(x) = \sin(\pi x)/(\pi x)$, $\mathrm{rect}(x) = 1$ for $|x| \le 1/2$ (0 otherwise). Next, we calculate the ensemble-averaged IPSF, $\langle \mathrm{IPSF}(x_i, y_i)\rangle$. Since

$$\langle \mathrm{IPSF}(x_i, y_i)\rangle = \left\langle \mathrm{APSF}(x_i, y_i)\mathrm{APSF}^*(x_i', y_i')\right\rangle\Big|_{x_i'=x_i,\; y_i'=y_i} \tag{S27}$$

it can be obtained from Eqs. (S24)–(S26) by setting $x_i' = x_i$ and $y_i' = y_i$, i.e.,

$$\langle \mathrm{IPSF}(x_i, y_i) \rangle = a^2 b^2 \,\mathrm{sinc}\left(\frac{bx_i}{\lambda f}\right) \mathrm{sinc}\left(\frac{by_i}{\lambda f}\right) \tag{S28}$$

Notably, when $b$ is sufficiently small,

$$\langle \mathrm{IPSF}(x_i, y_i) \rangle^2 = \langle \mathrm{IPSF} \rangle^2 = a^4 b^4 \tag{S29}$$

And

$$\Gamma(x_i, y_i, \Delta x_i, \Delta y_i) \approx \Gamma(\Delta x_i, \Delta y_i) = \langle \mathrm{IPSF} \rangle^2 [1 + \mathrm{sinc}^2\left(\frac{a\Delta x_i}{\lambda f}\right) \mathrm{sinc}^2\left(\frac{a\Delta y_i}{\lambda f}\right)] \tag{S30}$$

We can then obtain normalized autocovariance function (NAF) $C(\Delta x_i, \Delta y_i)$:

$$C(\Delta x_i, \Delta y_i) = \frac{\Gamma(\Delta x_i, \Delta y_i) - \langle \mathrm{IPSF} \rangle^2}{\langle \mathrm{IPSF} \rangle^2} = \mathrm{sinc}^2\left(\frac{a\Delta x_i}{\lambda f}\right) \mathrm{sinc}^2\left(\frac{a\Delta y_i}{\lambda f}\right) \tag{S31}$$

We further define the full width at half maximum (FWHM) of $C(\Delta x_i, \Delta y_i)$ as the average speckle size $SS_{\mathrm{avg}}$, then the average length of the speckle along the $x$ direction (which is the dispersion direction) is approximately

$$\mathrm{FWHM} \approx 0.886 \frac{\lambda f}{a} \tag{S32}$$

where $f$ is the focal length and $a$ is the size of the SGA. Note that this $SS_{\mathrm{avg}}$ is independent of optical system aberrations; this is an expected result due to the introduction of SGA. We prove in later notes that this innovation leads to a robust spectrometer that its performance is independent of the design, construction, and assembly errors of the spectrometer.

Refer to the schematic of the SHADES and consider $-1$ diffraction order. If there is a shift in wavelength $d\lambda$, the IPSF shifts an amount of $\frac{d\lambda f}{\Lambda}$. If this shift is greater than the $SS_{\mathrm{avg}}$, the spectrometer is then capable of resolving this wavelength shift, leading to a reasonable spectral resolution estimation at $\lambda_0$ as:

$$d\lambda_{\min} \approx 0.886 \frac{\lambda_0 \Lambda}{a} \tag{S33}$$

## Supplementary Note 3. SHADES with a multi-slit entrance aperture: fully uncorrelated PSFs

Next, we consider the geometry of the entrance aperture and its impact on the spectrometer's resolution. In this case, we model a plural of point sources $(x_j, y_j)$ on the entrance plane, their respective IPSFs superimpose to form a speckle pattern $I(x_i, y_i; \lambda)$ for a wavelength $\lambda$:

$$I(x_i,y_i;\lambda) = \sum_{j=1}^{N} \mathrm{IPSF}_j(x_i,y_i) * \delta\left(x_i - \left(\frac{m\lambda f}{\Lambda} - x_j\right), y_i + y_j\right) \tag{S34}$$

Following the definition in the previous section, we use the $SS_{\mathrm{avg}}$ (FWHM of the NAF) of $I(x_i,y_i)$ along the $x$-axis direction to estimate the spectral resolution. As the first case, let us first assume that $\mathrm{IPSF}_1$, $\mathrm{IPSF}_2$, $\mathrm{IPSF}_3$, ... are uncorrelated random processes. We note that this is an idealized limiting case; in practice, IPSFs generated by the same SGA are generally partially correlated. Here, however, we will consider this extreme case first. Assuming spatial stationarity over the region of interest:

$$\langle \mathrm{IPSF}_j(x_i,y_i)\rangle = \langle \mathrm{IPSF}(x_i,y_i)\rangle,\ \forall j \tag{S35}$$

Assuming $b$ is sufficiently small or $x_i$ and $y_i$ are small (i.e. central region of the PSF speckle), following Eq. (S28) (S29), ⟨IPSF(xi,yi)⟩ is a constant ⟨IPSF⟩ and it follows that

$$\langle I\rangle = I(x_i,y_i,\lambda) = N\cdot\langle \mathrm{IPSF}\rangle \tag{S36}$$

Thus,

$$I(x_i,y_i;\lambda) - \langle I\rangle = \sum_{j=1}^{N}\left[\mathrm{IPSF}_j\left(x_i + x_j - \frac{m\lambda f}{\Lambda}, y_i + y_j\right) - \langle \mathrm{IPSF}\rangle\right] \tag{S37}$$

We evaluate the covariance between the zero-mean intensity fluctuations contributed by two image plane points $(x_i,y_i)$ and $(x_i',y_i')$:

$$\langle (I(x_i,y_i;\lambda) - \langle I\rangle)(I(x_i',y_i';\lambda) - \langle I\rangle)\rangle \tag{S38}$$

Under the uncorrelated assumption, if $j \neq k$, the expectation of the product of their zero-mean fluctuations factorizes and vanishes:

$$\left\langle \left[\mathrm{IPSF}_j\left(x_i + x_j - \frac{m\lambda f}{\Lambda}, y_i + y_j\right) - \langle \mathrm{IPSF}\rangle\right]\left[\mathrm{IPSF}_k\left(x_i' + x_k - \frac{m\lambda f}{\Lambda}, y_i' + y_k\right) - \langle \mathrm{IPSF}\rangle\right]\right\rangle$$
$$= \begin{cases} \Gamma(\Delta x_i,\Delta y_i) - \langle \mathrm{IPSF}\rangle^2, & \text{if } j = k \\ 0, & \text{if } j \neq k \text{ (i.e. independent random processes)} \end{cases} \tag{S39}$$

where $\Delta x_i = x_i' - x_i$, $\Delta y_i = y_i' - y_i$. Therefore,

$$\langle (I(x_i,y_i;\lambda) - \langle I\rangle)(I(x_i',y_i';\lambda) - \langle I\rangle)\rangle = N[\Gamma(\Delta x_i,\Delta y_i) - \langle \mathrm{IPSF}\rangle^2] \tag{S40}$$

With Eq. (S30),

$$\Gamma(\Delta x_i, \Delta y_i) = \langle \text{IPSF} \rangle^2 \left[1 + \text{sinc}^2\left(\frac{a\Delta x_i}{\lambda f}\right) \text{sinc}^2\left(\frac{a\Delta y_i}{\lambda f}\right)\right] \tag{S41}$$

We can get

$$\langle (I(x_i, y_i; \lambda) - \langle I \rangle)(I(x_i', y_i'; \lambda) - \langle I \rangle) \rangle = N \cdot \langle \text{IPSF} \rangle^2 \, \text{sinc}^2\left(\frac{a\Delta x_i}{\lambda f}\right) \text{sinc}^2\left(\frac{a\Delta y_i}{\lambda f}\right) \tag{S42}$$

We can then obtain NAF for $I(x_i, y_i; \lambda)$, $C(\Delta x_i, \Delta y_i)$ as:

$$C(\Delta x_i, \Delta y_i) = \frac{\langle (I(x_i, y_i; \lambda) - \langle I \rangle)(I(x_i', y_i'; \lambda) - \langle I \rangle) \rangle}{N \langle PSF \rangle^2} = \text{sinc}^2\left(\frac{a\Delta x_i}{\lambda f}\right) \text{sinc}^2\left(\frac{a\Delta y_i}{\lambda f}\right) \tag{S43}$$

From the above equation, we can see that under the extreme assumption of uncorrelated intensity point spread functions, the $SS_{\text{avg}}$ is not changed despite a plural of point sources on the entrance aperture of the spectrometer. Hence, the resolution of the spectrometer is independent of the shape of the entrance aperture.

## Supplementary Note 4. SHADES with a multi-slit entrance aperture: fully correlated PSFs

We then consider the other side of the spectrum where IPSF is fully correlated, i.e. IPSF is shift-invariant. For a single wavelength, the intensity pattern on the image plane is the convolution between the IPSF and the image of entrance aperture function for a perfect spectrometer with no aberrations $E(x_i, y_i)$:

$$I(x_i, y_i; \lambda) = \text{IPSF}(x_i, y_i) * E(x_i, y_i) \tag{S44}$$

First, we calculate the ensemble mean of $I(x_i, y_i; \lambda)$:

$$\begin{aligned} \langle I \rangle = \langle I(x_i, y_i; \lambda) \rangle &= \iint \langle \text{IPSF}(x_i - \xi, y_i - \eta) \rangle E(\xi, \eta) \, d\xi \, d\eta \\ &= \iint \langle \text{IPSF} \rangle E(\xi, \eta) \, d\xi \, d\eta \end{aligned} \tag{S45}$$

Merging Eq. (S44) and Eq. (S45):

$$I(x_i, y_i; \lambda) - \langle I \rangle = [\text{IPSF}(x_i, y_i) - \langle \text{IPSF} \rangle] * E(x_i, y_i) \tag{S46}$$

Thus, the autocovariance can be computed as:

$$\begin{aligned}
&\Big\langle \big(I(x_i,y_i;\lambda)-\langle I\rangle\big)\big(I(x_i',y_i';\lambda)-\langle I\rangle\big)\Big\rangle \\
&\quad=\iiiint \Big\langle \big[\mathrm{IPSF}(x_i-\xi,\,y_i-\eta)-\langle\mathrm{IPSF}\rangle\big]\,\big[\mathrm{IPSF}(x_i'-\xi',\,y_i'-\eta')-\langle\mathrm{IPSF}\rangle\big]\Big\rangle \\
&\quad\quad\cdot E(\xi,\eta)\,E(\xi',\eta')\,d\xi\,d\eta\,d\xi'\,d\eta'.
\end{aligned} \tag{S47}$$

With $\Delta\xi=\xi'-\xi$, $\Delta\eta=\eta'-\eta$, $\Delta x_i=x_i'-x_i$ and $\Delta y_i=y_i'-y_i$ we obtain:

$$\begin{aligned}
&\Big\langle \big[\mathrm{IPSF}(x_i-\xi,\,y_i-\eta)-\langle\mathrm{IPSF}\rangle\big]\,\big[\mathrm{IPSF}(x_i'-\xi',\,y_i'-\eta')-\langle\mathrm{IPSF}\rangle\big]\Big\rangle \\
&\quad=\Gamma(\Delta x_i-\Delta\xi,\;\Delta y_i-\Delta\eta)-\langle\mathrm{IPSF}\rangle^2 \\
&\quad=\langle\mathrm{IPSF}\rangle^2\,\mathrm{sinc}^2\!\left(\frac{a(\Delta x_i-\Delta\xi)}{\lambda f}\right)\,\mathrm{sinc}^2\!\left(\frac{a(\Delta y_i-\Delta\eta)}{\lambda f}\right),
\end{aligned} \tag{S48}$$

Therefore,

$$\begin{aligned}
&\Big\langle \big(I(x_i,y_i;\lambda)-\langle I\rangle\big)\big(I(x_i',y_i';\lambda)-\langle I\rangle\big)\Big\rangle \\
&\quad=\iiiint \langle\mathrm{IPSF}\rangle^2\,\mathrm{sinc}^2\left(\frac{a(\Delta x_i-\Delta\xi)}{\lambda f}\right)\mathrm{sinc}^2\left(\frac{a(\Delta y_i-\Delta\eta)}{\lambda f}\right) \\
&\quad\quad\cdot E(\xi,\eta)E(\xi+\Delta\xi,\eta+\Delta\eta)\,d\xi\,d\eta\,d\Delta\xi\,d\Delta\eta
\end{aligned} \tag{S49}$$

Since $E$ is a real function,

$$\iint E(\xi,\eta)\,E(\xi+\Delta\xi,\eta+\Delta\eta)\,d\xi\,d\eta=(E\star E)(\Delta\xi,\Delta\eta) \tag{S50}$$

where $\star$ is the correlation operator, and this is the autocorrelation of $E$. Hence,

$$\begin{aligned}
\Big\langle \big(I(x_i,y_i;\lambda)-\langle I\rangle\big)\big(I(x_i',y_i';\lambda)-\langle I\rangle\big)\Big\rangle &= \langle\mathrm{IPSF}\rangle^2\left[\mathrm{sinc}^2\!\left(\frac{a\Delta x_i}{\lambda f}\right)\mathrm{sinc}^2\!\left(\frac{a\Delta y_i}{\lambda f}\right)\right] \\
&\quad * (E\star E)(\Delta x_i,\Delta y_i).
\end{aligned} \tag{S51}$$

We can then obtain the NAF in this case $C(\Delta x_i,\Delta y_i)$ as:

$$\begin{aligned}
C(\Delta x_i,\Delta y_i) &= \frac{\langle(I(x_i,y_i;\lambda)-\langle I\rangle)(I(x_i',y_i';\lambda)-\langle I\rangle)\rangle}{\langle\mathrm{IPSF}\rangle^2[(E\star E)(0,0)]} \\
&= \frac{\mathrm{sinc}^2\left(\frac{a\Delta x_i}{\lambda f}\right)\mathrm{sinc}^2\left(\frac{a\Delta y_i}{\lambda f}\right) * (E\star E)(\Delta x_i,\Delta y_i)}{\iint E^2(\xi,\eta)\,d\xi\,d\eta}
\end{aligned} \tag{S52}$$

It is interesting to note that this normalized autocorrelation is closely related to the incoherent optical transfer function in form. Defining

$$H(\Delta x_i, \Delta y_i) = \frac{(E \star E)(\Delta x_i, \Delta y_i)}{\iint E^2(\xi, \eta)\, d\xi\, d\eta} = \frac{(E \star E)(\Delta x_i, \Delta y_i)}{\iint E(\xi, \eta)\, d\xi\, d\eta} \tag{S53}$$

where, in the denominator, the fact that $E(\xi, \eta)$ is either 1 or 0 has been used to replace $E^2(\xi, \eta)$ by $E(\xi, \eta)$. Hence,

$$C(\Delta x_i, \Delta y_i) = \mathrm{sinc}^2\left(\frac{a\Delta x_i}{\lambda f}\right) \mathrm{sinc}^2\left(\frac{a\Delta y_i}{\lambda f}\right) * H(\Delta x_i, \Delta y_i) \tag{S54}$$

Then we evaluate entrance apertures of different shape. We first consider a slit on the entrance aperture plane, the slit width is $w_s$ and is along the $x_s$ direction, the slit length $l_s$ is along the $y_s$ direction. For slit aperture,

$$H(\Delta x_i, \Delta y_i) = \begin{cases} \frac{(w_s - |\Delta x_i|)(l_s - |\Delta y_i|)}{w_s l_s}, & |\Delta x_i| \leq w_s, |\Delta y_i| \leq l_s \\ 0, & \text{otherwise} \end{cases} \tag{S55}$$

Equivalently,

$$H(\Delta x_i, \Delta y_i) = \mathrm{tri}\left(\frac{\Delta x_i}{w_s}\right) \mathrm{tri}\left(\frac{\Delta y_i}{l_s}\right) \tag{S56}$$

Hence, the NAF can be computed as:

$$C(\Delta x_i, \Delta y_i) = \mathrm{sinc}^2\left(\frac{a\Delta x_i}{\lambda f}\right) \mathrm{sinc}^2\left(\frac{a\Delta y_i}{\lambda f}\right) * \left[\mathrm{tri}\left(\frac{\Delta x_i}{w_s}\right) \mathrm{tri}\left(\frac{\Delta y_i}{l_s}\right)\right] \tag{S57}$$

Again, we will use the FWHM of $C$ as a measure of the $SS_{\text{avg}}$. For the spectrometer application, as we only care about spectral resolution, we use $SS_{\text{avg}}$ along the $x_s$ axis and $C$ is now:

$$C(\Delta x_i) = \mathrm{sinc}^2\left(\frac{a\Delta x_i}{\lambda f}\right) * \mathrm{tri}\left(\frac{\Delta x_i}{w_s}\right) \tag{S58}$$

Therefore, the spectral resolution of SHADES with a slit opening and under the assumption of spatially invariant IPSF at $\lambda_0$ is established as:

$$\frac{d\lambda_{\min} f}{\Lambda} \approx 0.886 \frac{\lambda_0 f}{a} + w_s \quad \Rightarrow \quad d\lambda_{\min} \approx 0.886 \frac{\lambda_0 \Lambda}{a} + \frac{w_s \Lambda}{f} \tag{S59}$$

As we can see from above, if we use a single slit, SHADES suffers from the same trade-off between optical throughput and spectral resolution just like a traditional grating spectrometer.

Next, we consider an encoded multi-slit entrance aperture. The entrance aperture contains $M$ discrete slit positions, each with a slit width of $w_s$, and is encoded using a pseudo-random binary code. For such a code, our numerical analysis shows that the autocorrelation decreases to approximately 0.5 after a one-slit shift and remains below or around 0.5 for larger shifts. Therefore, when $M$ is large, the autocorrelation term $H(\Delta x_i)$ can be approximated as

$$H(\Delta x_i) \approx \frac{1}{2}\left[\mathrm{tri}\left(\frac{\Delta x_i}{w_s}\right) + \mathrm{tri}\left(\frac{\Delta x_i}{M w_s}\right)\right] \tag{S60}$$

where the first triangular term represents the narrow correlation feature associated with a single slit width, while the second triangular term describes the slowly decreasing base value determined by the full encoded aperture width. Since $M$ is large, the second term varies slowly near the origin, and $H(\Delta x_i)$ drops to approximately 0.5 at a displacement close to one slit width. Therefore, the effective FWHM of $H(\Delta x_i)$ is approximately $2w_s$.

Then, with the convolution in Eq. S50, the FWHM of $C$ is mainly determined by $\mathrm{FWHM}(C) \approx 0.886\frac{\lambda f}{a} + 2w_s$. This indicates that the spectral resolution is still primarily governed by the diffraction-limited response and a single slit width in the array, rather than by the full encoded aperture size. At the same time, because approximately half of the $M$ slit positions are open in the pseudo-random binary code, the optical throughput is enhanced by roughly $M/2$ times compared with a single-slit aperture. Equivalently, the effective light-collecting area becomes approximately $M w_s l_s/2$, where $l_s$ is the slit length. Therefore, the encoded multi-slit aperture greatly increases photon throughput while preserving a resolution scale set by the smallest slit width, thereby relaxing the conventional trade-off between spectral resolution and optical throughput.

The numerical results in Fig. S2 provide a direct interpretation of the practical resolution scale of the encoded aperture. In this simulation, the aperture pattern is laterally shifted by different amounts and added to the original pattern, forming $A_{\mathrm{sum}} = A + A_{\mathrm{shift}}$. The correlation between this superimposed pattern and the original aperture, $\mathrm{corr}(A_{\mathrm{sum}}, A)$, is then evaluated. This procedure mimics the situation in which two neighboring spectral components generate two shifted encoded responses on the detector, and the computational reconstruction attempts to distinguish them through correlation with the calibrated aperture response.

For sub-slit displacements, the two shifted responses are too close to each other, so their correlation contributions strongly overlap and merge into a single broadened response. In this regime, the superimposed pattern $A_{\mathrm{sum}}$ remains highly similar to the original pattern, and the two spectral components are therefore largely indistinguishable. Once the displacement reaches approximately one slit width, the correlation response begins to broaden and decorrelate noticeably, indicating the onset of resolvability. However, this should be interpreted as the beginning of separation rather than a conservative resolution limit, because the two correlation features still retain substantial overlap and may be sensitive to noise, calibration error, and reconstruction uncertainty.

As the displacement further increases toward two slit widths, the two correlation contributions in $\mathrm{corr}(A_{\mathrm{sum}}, A)$ become more clearly separated. This corresponds to a more robust criterion for distinguishing two adjacent spectral components in the computational reconstruction. Therefore, the practical computational resolution should be regarded as a transition range rather than a single sharp cutoff. In this work, we estimate the effective resolution to lie between one and two slit widths, namely $w_s$–$2w_s$, where $w_s$ represents the onset of distinguishability and $2w_s$ represents a conservative FWHM-based resolvability estimate.

## Supplementary Note 5. SHADES with a multi-slit entrance aperture: partially correlated PSFs

In practice, SHADES typically operates in a partially correlated regime. Therefore, we evaluate the mutual correlation between IPSFs produced by two different entrance points. This two-point IPSF correlation provides a direct measure whether the system behavior is closer to the fully correlated or fully uncorrelated limit. We consider the correlation between two IPSFs from two different point sources located on the entrance aperture plane

$$\Gamma_{12}(x_i,y_i;x_i',y_i') = \langle \mathrm{IPSF}_1(x_i,y_i)\mathrm{IPSF}_2(x_i',y_i')\rangle \tag{S61}$$

$\mathrm{IPSF}_1(x_i,y_i)$ is the IPSF from a point $(x_1,y_1)$ and $\mathrm{IPSF}_2(x_i,y_i)$ is from another point $(x_2,y_2)$ on the entrance plane.

$$\Gamma_{12}(x_i,y_i;x_i',y_i') = \langle \mathrm{IPSF}(x_i,y_i)\rangle^2 \left[1+\frac{|\langle \mathrm{APSF}_1(x_i,y_i)\mathrm{APSF}_2^*(x_i',y_i')\rangle|}{\langle \mathrm{IPSF}(x_i,y_i)\rangle^2}^2\right] \tag{S62}$$

The above used the fact that $\langle \mathrm{IPSF}_1(x_i,y_i)\rangle = \langle \mathrm{IPSF}_2(x_i,y_i)\rangle = \langle \mathrm{IPSF}(x_i,y_i)\rangle$. Similarly, we have

$$\begin{aligned}\langle \mathrm{APSF}_1(x_i,y_i)\mathrm{APSF}_2^*(x_i',y_i')\rangle = \iiiint &\langle P_1(x_g,y_g)P_2^*(x_g',y_g')\rangle \\ &\cdot e^{-i\frac{2\pi}{\lambda f}(x_ix_g+y_iy_g)}e^{i\frac{2\pi}{\lambda f}(x_i'x_g'+y_i'y_g')}\,dx_g\,dy_g\,dx_g'\,dy_g'\end{aligned} \tag{S63}$$

With

$$\begin{aligned}P_1(x_g,y_g) &= G(x_g,y_g)e^{i[kW_1(x_g,y_g)+\phi_r(x_g,y_g)]}\\ P_2(x_g',y_g') &= G(x_g',y_g')e^{i[kW_2(x_g',y_g')+\phi_r(x_g',y_g')]}\end{aligned} \tag{S64}$$

$$\langle P_1(x_g,y_g)P_2^*(x_g',y_g')\rangle = G(x_g,y_g)e^{ik[W_1(x_g,y_g)-W_2(x_g,y_g)]}\mu(\Delta x_g,\Delta y_g) \tag{S65}$$

Define the difference in wavefront error

$$\Delta\phi(x_g,y_g) = k[W_1(x_g,y_g)-W_2(x_g,y_g)] \tag{S66}$$

$$\begin{aligned}\langle \mathrm{APSF}_1(x_i,y_i)\,\mathrm{APSF}_2^*(x_i+\Delta x_i,y_i+\Delta y_i)\rangle = &\iint G(x_g,y_g)\,e^{-i\frac{2\pi}{\lambda f}(\Delta x_ix_g+\Delta y_iy_g)}\,e^{i\Delta\phi(x_g,y_g)}\,dx_g\,dy_g\\ &\cdot\iint \mu(\Delta x_g,\Delta y_g)\,e^{-i\frac{2\pi}{\lambda f}(x_i\Delta x_g+y_i\Delta y_g)}\,d\Delta x_g\,d\Delta y_g\end{aligned} \tag{S67}$$

If we assume a very small micro grating cell size or are only interested in the central region of the speckle,

$$\iint \mu(\Delta x_g, \Delta y_g) e^{-i\frac{2\pi}{\lambda f}(x_i \Delta x_g + y_i \Delta y_g)}\, d\Delta x_g\, d\Delta y_g = \text{constant} \tag{S68}$$

And, setting $\text{APSF}_1$ and $\text{APSF}_2$ to APSF, $\Delta x_i = 0$, and $\Delta y_i = 0$ in Eq. S67, we can obtain:

$$\langle \text{IPSF}(x_i, y_i) \rangle^2 = \langle \text{IPSF} \rangle = \text{constant} \iint G(x_g, y_g) dx_g dy_g \tag{S69}$$

Inserting Eq. S67, S68, S69 into S62, and more specifically, we are interested in the correlation of $\text{IPSF}_1$ and $\text{IPSF}_2$ when $\Delta x_i = 0$ and $\Delta y_i = 0$. Hence, we have

$$\Gamma_{12}(0,0) = \langle \text{IPSF} \rangle^2 \left[ 1 + \left| \frac{\iint G(x_g, y_g) e^{i\Delta\phi(x_g, y_g)}\, dx_g\, dy_g}{\iint G(x_g, y_g)\, dx_g\, dy_g} \right|^2 \right] \tag{S70}$$

Similarly, the NAF is:

$$C_{12}(0,0) = \left| \frac{\iint G(x_g, y_g) e^{i\Delta\phi(x_g, y_g)}\, dx_g\, dy_g}{\iint G(x_g, y_g)\, dx_g\, dy_g} \right|^2 \tag{S71}$$

Clearly, when the difference between the two wavefront errors is zero, despite the amount of each individual wavefront error, i.e. $\Delta\phi(x_g, y_g) = 0$, $C_{12}(0,0) = 1$, the two IPSFs are identical. However, when $\Delta\phi(x_g, y_g)$ increases, the $C_{12}(0,0)$ decreases, i.e. the correlation between two IPSFs decreases. Based on the deduction above, we can summarize as below. If the IPSFs are fully uncorrelated, then the $SS_{\text{avg}}$ in a single wavelength response pattern of the spectrometer is independent of the shape of the entrance aperture, and it is equal to the $SS_{\text{avg}}$ in a single IPSF. If the IPSFs are fully correlated, i.e. spatially invariant, then the $SS_{\text{avg}}$ in a single wavelength response pattern of the spectrometer is equal to the $SS_{\text{avg}}$ in a single IPSF broadened by a single slit width of the entrance aperture.

More interestingly, if we use a pseudo-random binary array encoded multi-slit aperture at the entrance, then SHADES has the ability to decouple the spectral resolution from its optical throughput, resulting in a spectrometer both free of construct errors and high optical throughput. In practice, the IPSFs from different points on the entrance aperture plane are neither fully correlated nor completely uncorrelated. Their correlations are dependent on the differences between wavefront errors on the SGA plane. However, for a conservative estimate of SHADES spectral resolution, we can use the spatially invariant assumption of IPSFs and design the SHADES accordingly.

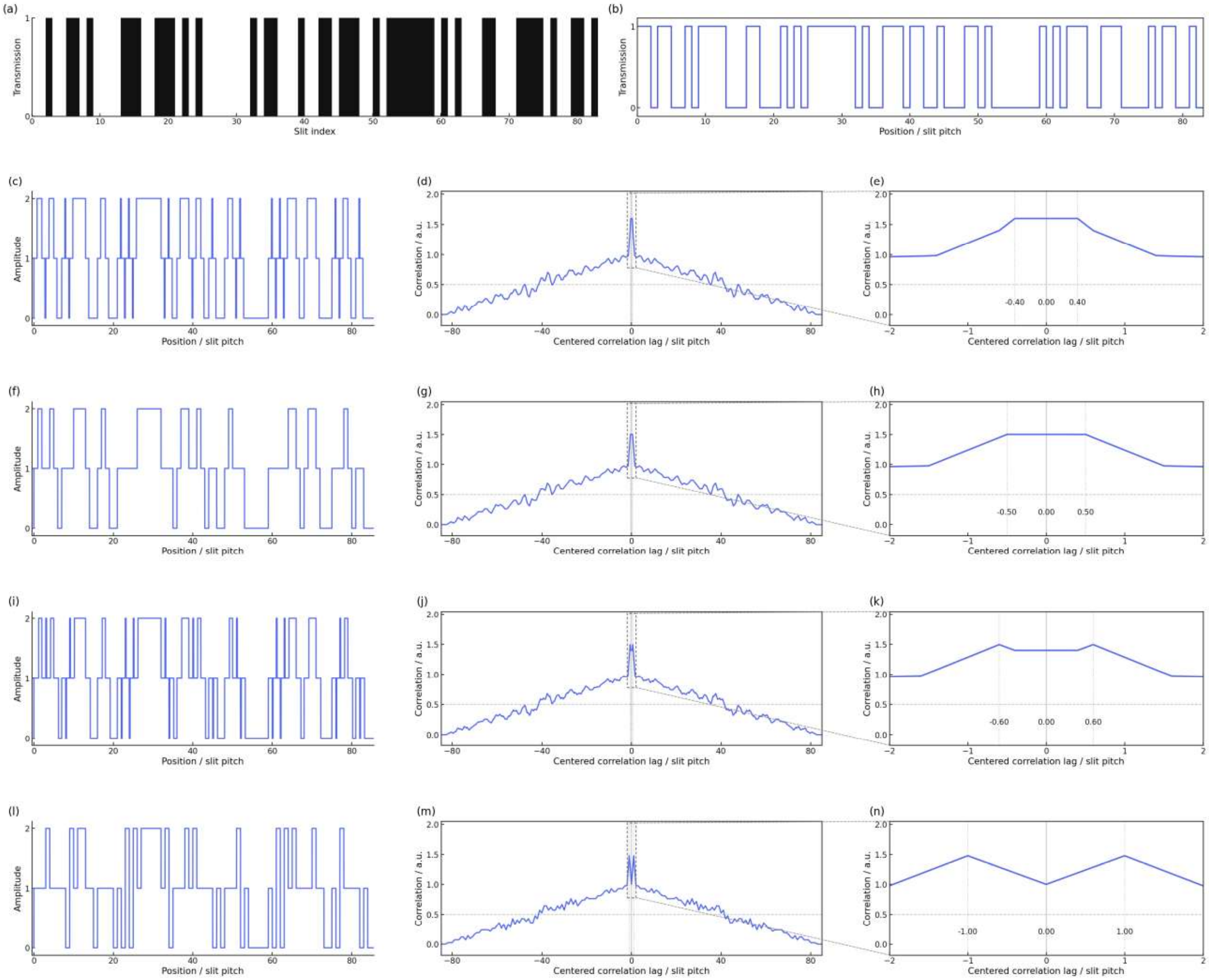

**Fig. S2. Resolution analysis of the encoded entrance aperture with a total aperture size of 83 slit pitches.** (a) Binary transmission pattern of the encoded entrance aperture, plotted as transmission versus slit index. The aperture contains 83 slit-pitch positions, where the white bars denote transmissive elements and black bars denote blocked elements. (b) One-dimensional transmission profile of the same encoded aperture, plotted as transmission versus position normalized by the slit pitch. (c, f, i, l) Superposed aperture responses formed by two identical encoded patterns with center-to-center separations of 0.8, 1.0, 1.2, and 2.0 slit pitches, respectively. (d, g, j, m) Corresponding correlation curves between the original encoded pattern and the superposed patterns in (c, f, i, l), shown over the full centered lag range. (e, h, k, n) Enlarged views of the central correlation regions in (d, g, j, m). The dashed vertical lines indicate the expected positions of the two shifted responses for center-to-center separations of 0.8, 1.0, 1.2, and 2.0 slit pitches, respectively. As the separation increases, the two correlation features become progressively more distinguishable. This comparison indicates that the encoded entrance aperture with an 83-pitch size provides a practical resolution between one and two slit pitches.

## Supplementary Note 6. Optical design of μSHADES

As shown in Fig. S3a–b, μSHADES adopts a compact off-axis Ebert optical layout. A concave mirror with a focal length of 4.5 mm collimates the beam from the coded multi-slit aperture, which illuminates the SGA. The dispersed wavefront is then imaged by the same concave mirror onto the detector with an effective f/# ≈ 3, forming the speckle-encoded spectral response on the CMOS sensor. This folded architecture improves packaging density; however, off-axis aberrations are significant, as shown in Fig. S3c.

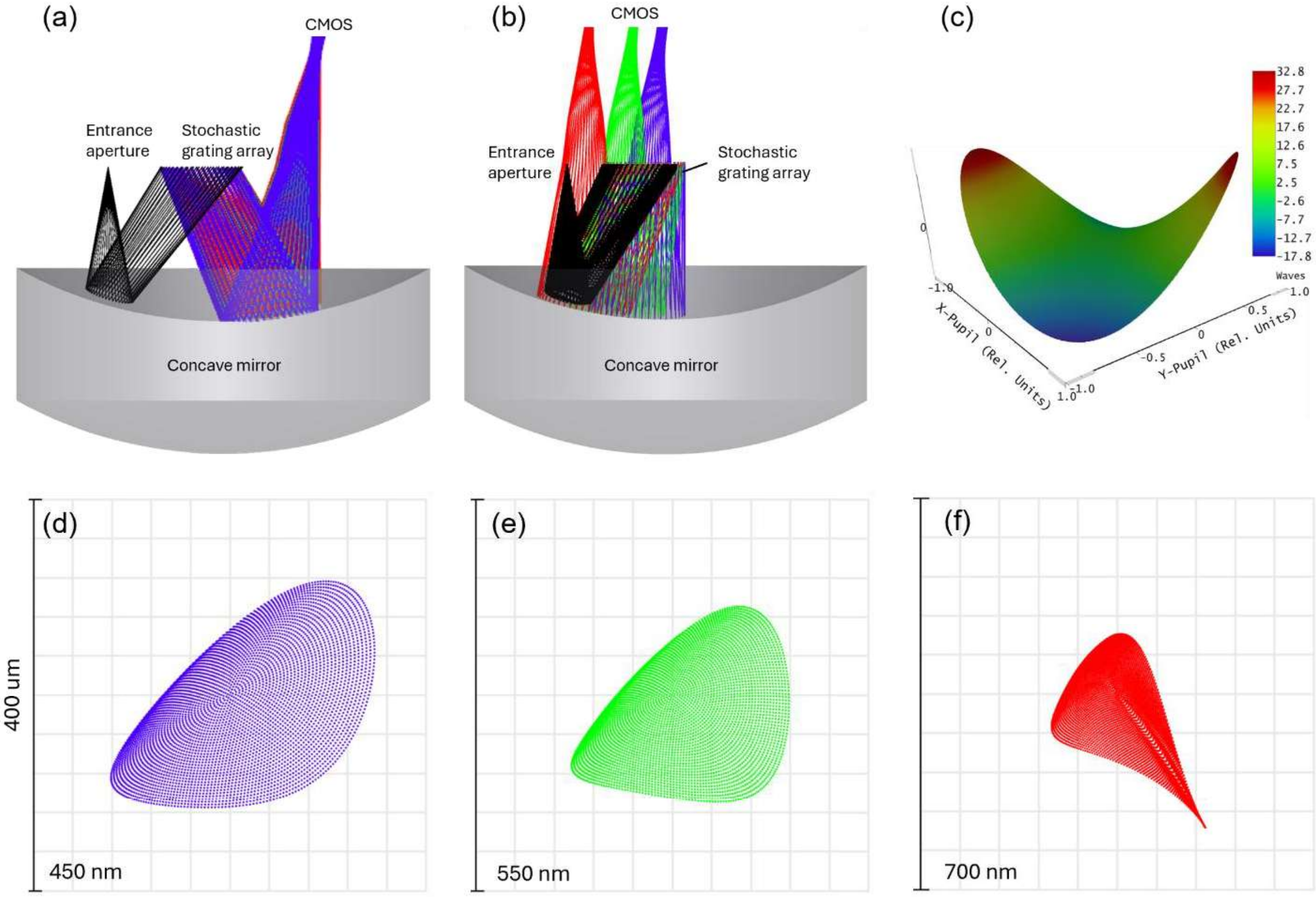

**Fig. S3. Ray-tracing model and aberration analysis of μSHADES.** (a,b) 3D ray-tracing model of the folded optical architecture shown from the side and the top, illustrating the folded beam path between the entrance aperture, concave mirror, the SGA, and the CMOS detector plane. (c) Aberrated pupil-plane wavefront error exported from Zemax at 550 nm, plotted in units of waves; pupil coordinates are normalized. (d–f) Simulated geometric spot diagrams at 450 nm, 550 nm, and 700 nm, respectively; scale displayed is 400 μm.

In the Zemax design, the SGA is replaced with a conventional reflective grating element (no stochastic phase). The resulting geometric spot diagrams at 450 nm, 550 nm, and 700 nm are shown in Fig. S3d–f. As the wavelength shifts away from the center, the spots become increasingly elongated and skewed, indicating stronger wavelength-dependent and field-dependent distortions. This trend is consistent with the combined effects of wavelength-dependent and off-axis aberrations inherent to the folded geometry. Together, the wavelength-dependent spot distortions and the aberration-limited baseline resolution constrain the native resolving power of μSHADES and motivate the stochastic-phase encoding and computational correction strategies described in the main text.

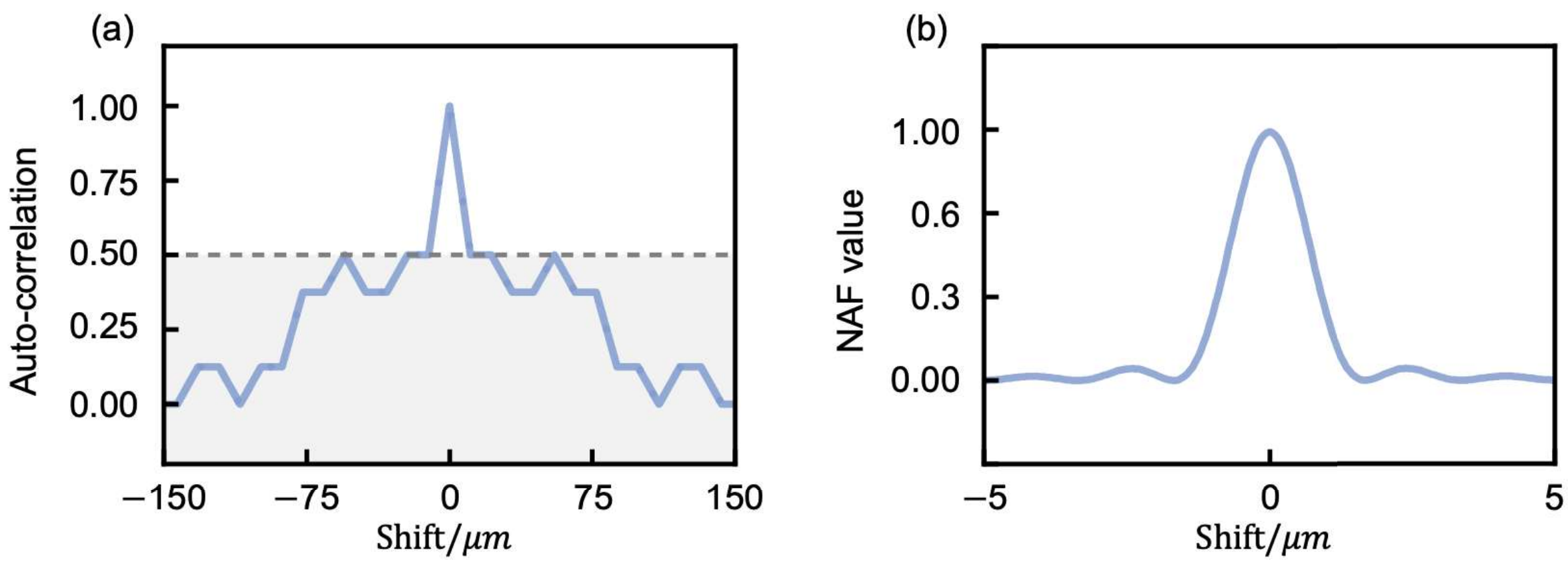

**Fig. S4. Response analysis of aperture and point.** (a) Autocorrelation of the aperture used in our experiment. (b) NAF of the point response on the entrance aperture.

Also, Fig. S4 shows the autocorrelation of the entrance aperture and the NAF of the point response associated with each entrance-aperture position. The autocorrelation of the entrance aperture decreases to 0.5 after a shift of one slit width and remains at or below 0.5 for larger shifts, as shown in Fig. S4a. We note that different entrance-aperture coding patterns may produce different autocorrelation functions, although most orthogonal coding schemes exhibit the same general behavior. The NAF of each point response follows a sinc function, as also shown in Fig. S4b. The overall response is therefore the convolution of these two functions, as described in the main text.

## Supplementary Note 7. Fabrication of SGA and multi-slit entrance aperture

The device is fabricated on a 500-μm-thick fused-silica ($SiO_2$) substrate using a two-level lithography and lift-off process to integrate the SGA with an encoded multi-slit entrance aperture, followed by a secondary metal masking layer for stray-light suppression and definition of a clear detector window. The process is shown in Fig. S5. In brief, the flow comprises two resist coating/exposure steps and two aluminum depositions (134 nm and 60 nm, respectively), each terminated by lift-off, enabling independent control of the sub-micrometer nanostructures and the larger-scale aperture/window features.

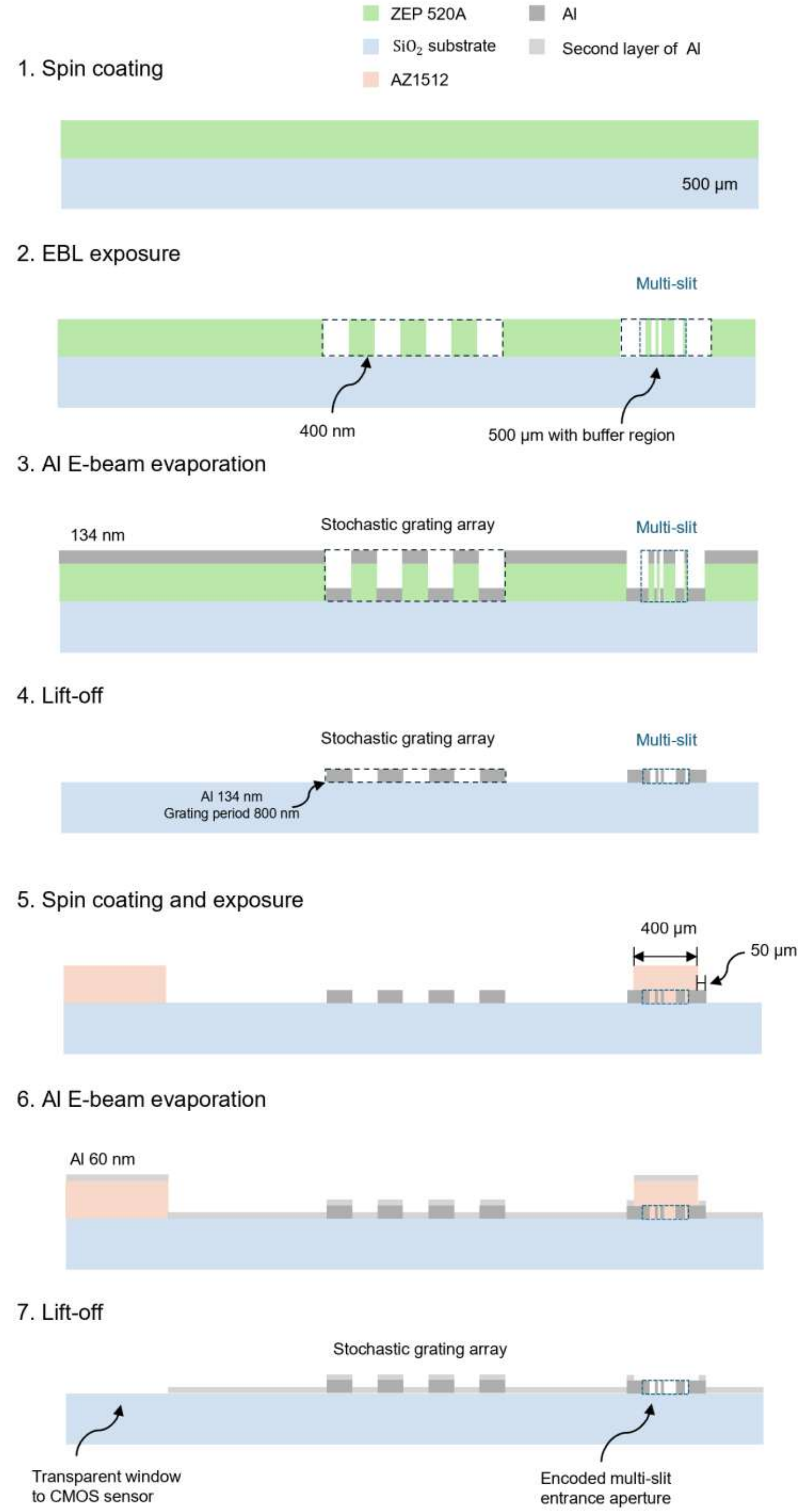

**Fig. S5. Microfabrication process flow for the integrated SGA and multi-slit entrance aperture.** The device is fabricated on a 500-μm-thick fused-silica ($\boldsymbol{SiO_2}$) substrate using two lithography-and-lift-off cycles with sequential Al depositions. (Steps 1–4) ZEP 520A e-beam resist is spin-coated (Step 1) and patterned by electron-beam lithography to define both the SGA region and the encoded multi-slit aperture field (Step 2), followed by Al deposition (134 nm) by e-beam evaporation (Step 3) and lift-off to form the first-level Al nanostructures (Step 4). (Steps 5–7) A second aligned lithography step defines a large-area optical blocking mask and the opening/window aligned to the CMOS sensor region (Step 5), followed by a second Al deposition (60 nm) (Step 6) and lift-off (Step 7) to form the final masking layer to block stray light while preserving optical access to the first-level SGA and the encoded multi-slit aperture.

For the first level of Al, ZEP 520A electron-beam resist is spin-coated onto the $SiO_2$ substrate (Step 1) and patterned by e-beam lithography to define both the SGA and the multi-slit region (Step 2). Alignment marks are patterned in the same lithography step to enable overlay of the second level. The SGA is designed with a period of 800 nm and a duty cycle of 50%, while the multi-slit aperture is written as a dedicated field with an adjacent buffer region to isolate the aperture from surrounding structures and to reduce edge-related patterning artifacts. After development, an Al film of 134 nm is deposited by electron-beam evaporation (Step 3). Lift-off removes the resist and excess metal, leaving the patterned Al structures that form the functional SGA and the encoded multi-slit aperture on the substrate (Step 4).

A second lithography step is then carried out to implement an optical blocking mask and to define a transparent window aligned to the CMOS sensor region. The second lithography level is registered to the first metal layer using alignment marks. A new resist layer is applied and patterned to form large-area metal masking, together with the aperture field (Step 5). A 50-μm buffer region is included to ensure overlap. Following pattern development, a second Al layer (60 nm) is deposited by e-beam evaporation (Step 6) and lifted off to form the final masking architecture (Step 7). The completed device, therefore, combines the first-level Al nanostructures (SGA and encoded multi-slit aperture; 134 nm) with a superimposed, large-area Al mask (60 nm) that suppresses parasitic illumination and provides a well-defined clear window for detector readout. Also, as shown in Fig. S6, s-polarized light shows lower efficiency than p-polarized light, along with a less uniform distribution over the functional range.

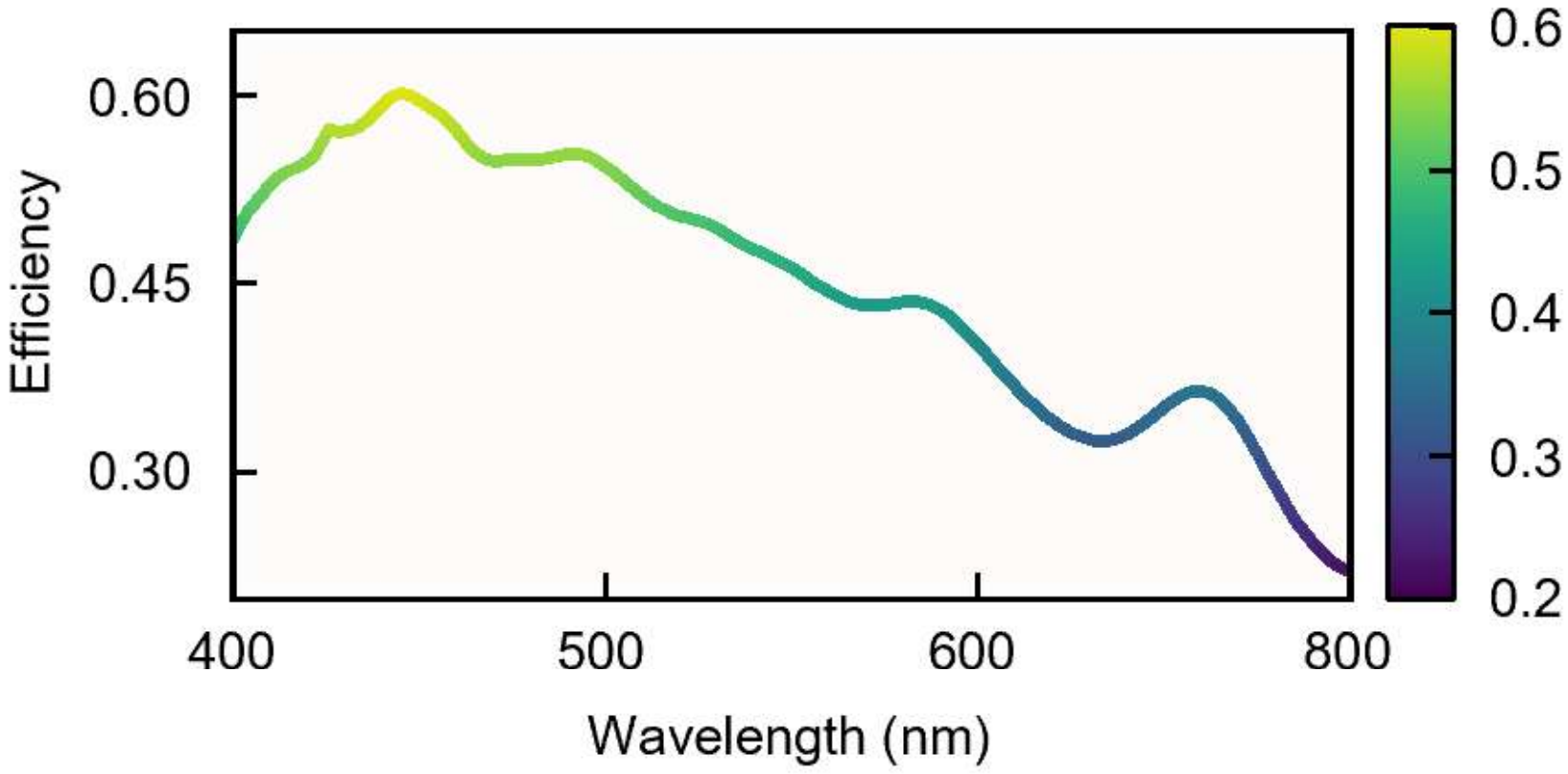

**Fig. S6. Efficacy of grating for s-polarized input light.**

## Supplementary Note 8. Optical design of the ASG

The optical layout of the arbitrary spectrum generator (ASG) is illustrated in Fig. S7. The system functions by spatially dispersing broadband light, modulating it via a programmable digital micromirror device (DMD), and recombining the selected components to synthesize a custom output spectrum. Emission from the source first passes through a slit (SL) placed at the object plane, which suppresses lateral spatial information and defines the input beam geometry. The beam is then redirected by a reflective mirror (RM) and collimated by a concave mirror (CM1) before reaching the first grating (G1). This grating disperses the collimated light horizontally along the spectral axis, mapping the wavelength distribution onto the surface of the DMD.

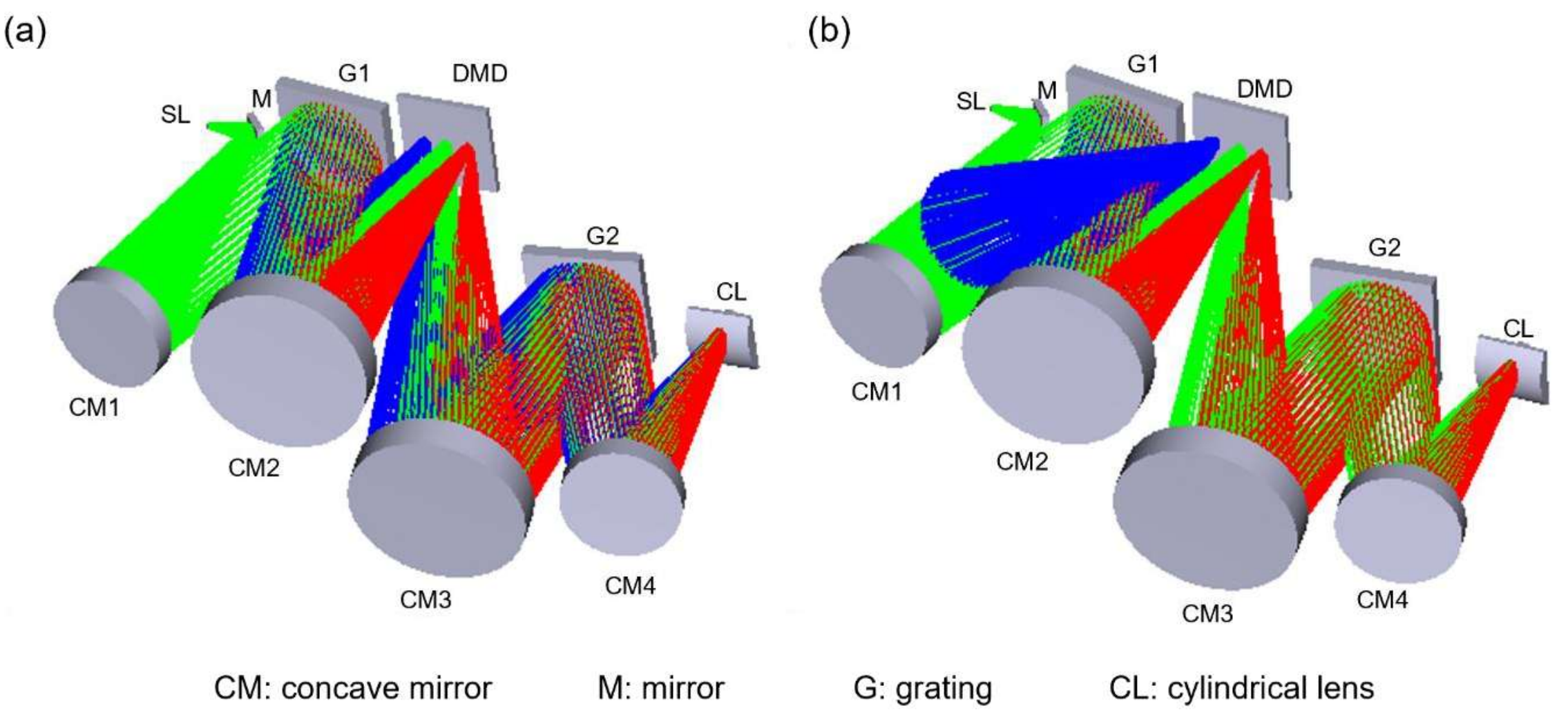

**Fig. S7. Optical architecture and ray-tracing simulation of the ASG.** (a) Spectral synthesis path (ON state). Light enters through the slit (SL) and is redirected by a mirror (M). Rays corresponding to the selected wavelength components are relayed by the concave mirrors (CM1–CM2) and spectrally dispersed by the first grating stage (G1), forming a spatially separated spectrum at the DMD plane. Micromirrors in the ON state redirect the selected wavelengths into the recombination arm, where the light is relayed by CM3 to the second grating stage (G2) for spectral recombination. The recombined beam is then directed by CM4 and reshaped by a cylindrical lens (CL) and coupled into the light guide, producing the synthesized target spectrum. (b) Spectral rejection path (OFF state). Unwanted wavelength components are steered by the DMD micromirrors in the OFF state away from the recombination path, preventing them from reaching the output. Together, the two paths illustrate how wavelength-selective routing on the DMD enables arbitrary spectral shaping.

At the modulation plane, the DMD operates as a binary spectral filter where each column corresponds to a distinct wavelength channel and the rows provide intensity control. Although the DMD active area comprises 1080 (rows) × 1920 (columns) pixels, we utilize a central region of 400 (rows) $\times$ 1160 (columns) pixels to cover the band of use. Micromirrors in the "ON" state reflect specific spectral components toward the second grating (G2). G2 is identical to G1 and operates in a reverse dispersion mode, recombining the spatially separated wavelengths into a single collimated beam, which is then reshaped by a cylindrical lens (CL) and then focused into the output light guide by another focusing lens. Conversely, micromirrors in the "OFF" state dump the unwanted wavelengths (as shown in Fig. S7b).

The performance of the ASG, particularly the achievable spectral resolution, is determined mainly by the entrance slit width, the grating groove density, and the DMD pixel pitch. Narrowing the input slit improves the spectral resolution but proportionally decreases optical throughput, reflecting the standard trade-off between spectral fidelity and intensity. The system is spectrally calibrated to map the DMD column indices to specific wavelengths, allowing for the precise, controlled generation of arbitrary spectral profiles.

## Supplementary Note 9. Calibration of the ASG spatial distribution

Although the ASG enables precise wavelength selection via the DMD columns, the incident illumination on the DMD is spatially non-uniform. Consequently, the optical power contributed by a specific wavelength channel is not linearly proportional to the simple summation of the "ON" micromirrors in that column. To ensure high-fidelity spectral shaping, we must calibrate this inhomogeneity to establish a precise mapping between the DMD pattern and the output intensity.

A straightforward raster scan, which turns on one pixel at a time, is too low in intensity to measure and brings about a low signal-to-noise ratio. Multiplexing of the whole DMD area will require a large-scale matrix inversion and an impractically long computation time. These will cause errors to spread, and it is less computationally practical. Instead, we adopt a column-wise multiplexing approach based on an S-matrix array. As shown in Fig. S8a, we sequentially step through the wavelength columns within the functional range. For each column, which represents a distinct wavelength, we resolve the vertical intensity profile by displaying a sequence of binary patterns from a cyclic S-matrix array of order 419, which is larger than the utilized vertical pixel count. The outer 19 rows are not used in subsequent spectrum synthesis.

Let $p \in R^N$ denotes the unknown vertical intensity profile for a given column. We acquire $N$ single-pixel measurements, denoted by the vector $m \in R^N$, by projecting the binary S-matrix array patterns. The S-matrix is inherently cyclic and full-rank and an illustrative example (order 7) is shown in Fig. S8b. The intensity profile is then reconstructed via linear inversion:

$$p = S^{-1}m \qquad (S72)$$

where $S$ is the $N \times N$ S-matrix. This process yields the experimental illumination map shown in Fig. S8c, which is subsequently used to generate the lookup tables for arbitrary spectrum synthesis.

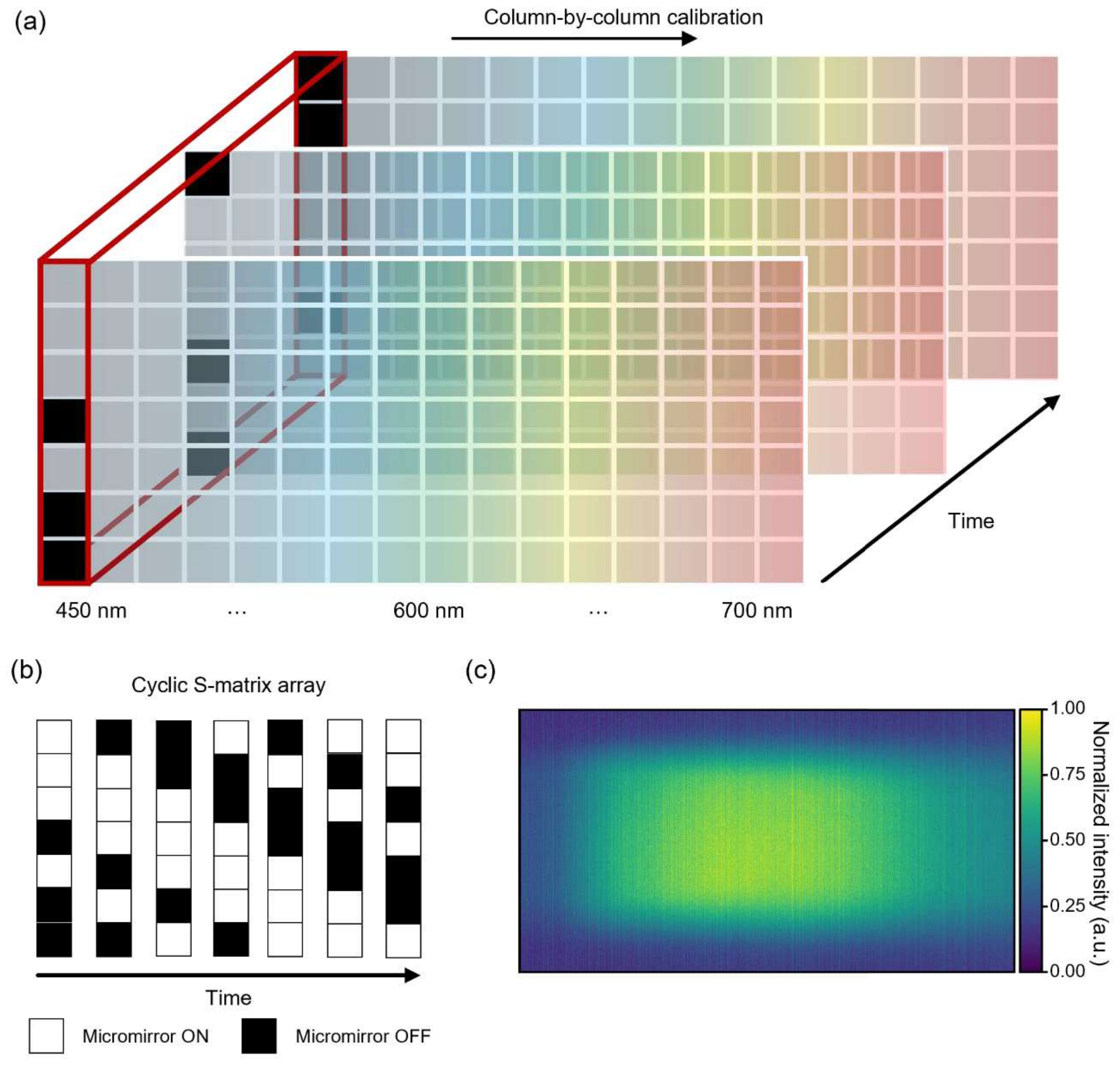

**Fig. S8. Calibration of the ASG spatial illumination on the DMD.** (a) Schematic of the column-wise calibration strategy. The DMD is segmented by wavelength columns. The calibration proceeds sequentially across the spectral range (indicated by the arrow) to characterize the intensity contribution of each pixel. (b) Representation of the cyclic S-matrix used for multiplexed measurement. A simplified order-7 S-matrix is shown here for illustration; the actual experiment utilizes an order-419 S-matrix to resolve the vertical intensity profile within each column. (c) Experimentally measured illumination map of the dispersed broadband illumination within the utilized DMD region, showing a non-uniform (approximately Gaussian-like) intensity distribution that is compensated for during spectral synthesis.

## Supplementary Note 10. ASG spectrum generation

With the per-column spatial illumination map established (as detailed in Supplementary Note 9), the spectral synthesis is formulated as a combinatorial optimization problem. For each wavelength channel (DMD column), the objective is to select a specific subset of micromirrors to toggle "ON" such that the integrated reflected intensity matches a target spectral density.

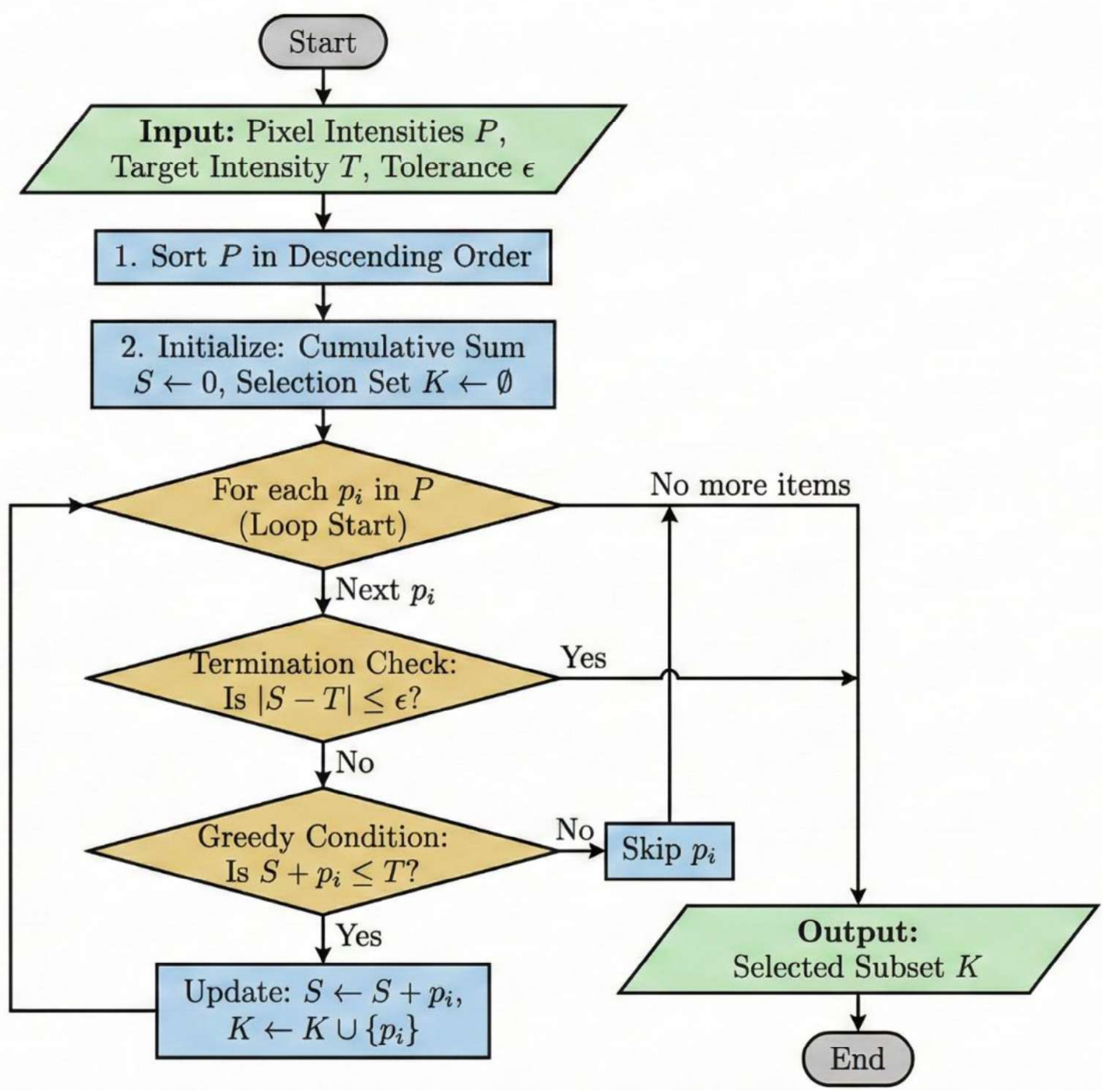

**Fig. S9. Flowchart of the GO algorithm for spectrum generation.** The GO algorithm selects a subset $\boldsymbol{K}$ of DMD pixels to be activated within a given wavelength column in order to approximate a desired target intensity $\boldsymbol{T}$ within a convergence tolerance $\boldsymbol{\varepsilon}$. Pixels are first sorted by intensity in descending order to maximize convergence speed. The algorithm iterates through the candidate pixels, accumulating intensity $\boldsymbol{S}$ only if the addition of a pixel keeps the total sum within the target limit, $\boldsymbol{S} + \boldsymbol{p_i} \leq \boldsymbol{T}$. The process terminates when the accumulated intensity approximates the target within the tolerance $\boldsymbol{\varepsilon}$ or when all pixels have been evaluated.

This formulation can be cast as a subset-sum-like selection problem. Classical solvers like dynamic programming [1] target exact or near-exact solutions and therefore, they are computationally prohibitive for real-time applications [2], especially given the high dimensionality of the DMD ($N = 400$ pixels per column). However, since the optical intensity is additive across DMD rows, a heuristic approach suffices. We therefore adopt a greedy optimization (GO) algorithm, which offers a computationally efficient solution with negligible synthesis error. The logic flow is detailed in Fig. S9.

The optimization routine processes each wavelength channel independently. Let $P = \{p_1, p_2, \dots, p_N\}$ represent the vector of calibrated pixel intensities for a given column and $T$ denote the target intensity. The procedure begins by sorting the elements of $P$ in descending order of magnitude. This prioritization ensures that pixels with the largest intensity contributions are evaluated first, accelerating convergence and reducing noise.

During the iterative phase, the algorithm maintains a cumulative sum $S$ (initialized to 0) and a selection set $K$. For each candidate pixel $p_i$ in the sorted sequence, a termination check is first performed: if the absolute error $|S - T|$ falls within a predefined tolerance $\varepsilon$, the iteration ceases immediately to minimize computation time. If convergence is not yet achieved, a greedy condition is evaluated; the pixel $p_i$ is added to the selection set $K$ if and only if its inclusion does not cause the cumulative sum to exceed the target (i.e., $S + p_i \leq T$). This heuristic approach rapidly constructs a near-optimal local solution, enabling fast spectral switching and synthesis.

## Supplementary Note 11. Experimental setup of the dataset capture

To train the deep learning (DL) reconstruction model, we constructed a dual-path optical characterization setup capable of acquiring synchronized measurements of the spectral ground truth and the corresponding μSHADES measurement. The optical design is shown in Fig. S10a, and the experimental layout is detailed in Fig. S10b.

As shown in Fig. S10a, light generated by the ASG is delivered via a light guide. The diverging output is collimated by lenses 1, 2 and 3 and split into two optical paths by a beam splitter. The reflected path is focused by lens 4 into a commercial reference spectrometer (Ocean Optics FLAME-T-VIS-NIR-ES), which records the high-fidelity spectral intensity profile serving as the ground truth label. Simultaneously, the transmitted path is focused by lens 5 onto the multi-slit entrance aperture of the μSHADES to capture the corresponding raw measurement. A three-axis translational stage is used to align the focused beam to the μSHADES entrance aperture.

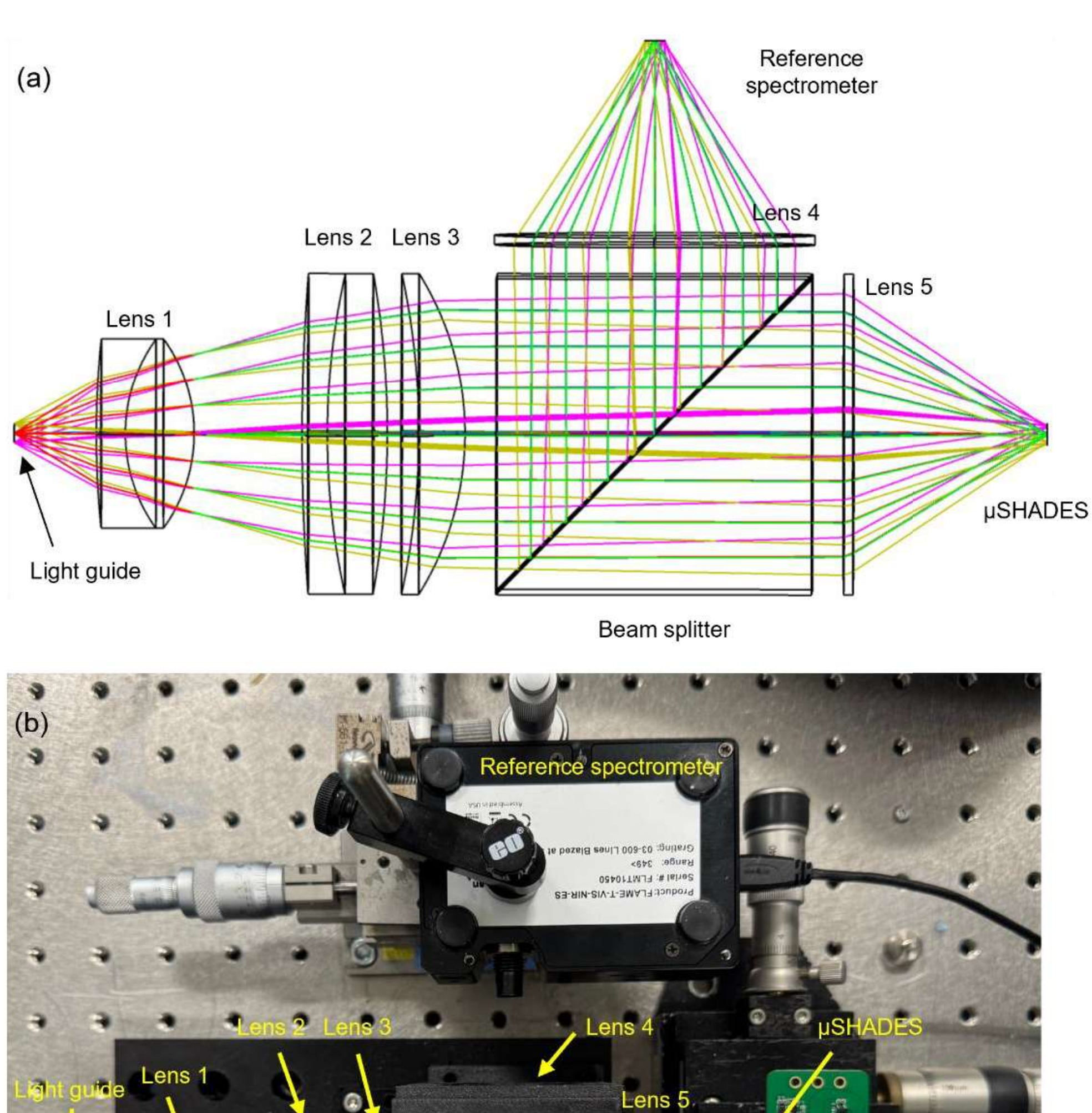

**Fig. S10. Experimental setup of µSHADES dataset capture.** (a) Ray-tracing schematic of the optical path. Light output from the guide is collimated and relayed through a series of lenses (Lens 1–3) before reaching a beam splitter, which splits the signal to µSHADES and a reference spectrometer. Lens 4 couples/focuses into the reference spectrometer; lens 5 focuses onto the µSHADES entrance aperture. (b) Photograph of the experimental setup. The system integrates the optical delivery components with a reference spectrometer mounted on one side of the beam splitter. This dual-path configuration allows for simultaneous acquisition of the ground truth spectrum and the µSHADES measurement.

## Supplementary Note 12. Evaluation of spectral reconstruction performance

To evaluate the generalization capability of the DL reconstruction model, μSHADES is challenged with a diverse set of test spectral profiles generated by the ASG. These profiles are strictly excluded from the training dataset to ensure independent validation. The test set comprises two distinct signal categories: sparse, narrowband signals mimicking atomic emission lines, and continuous broadband distributions representative of fluorescence emission and broadband illumination spectra. The comparative reconstruction results are summarized in Fig. S11.

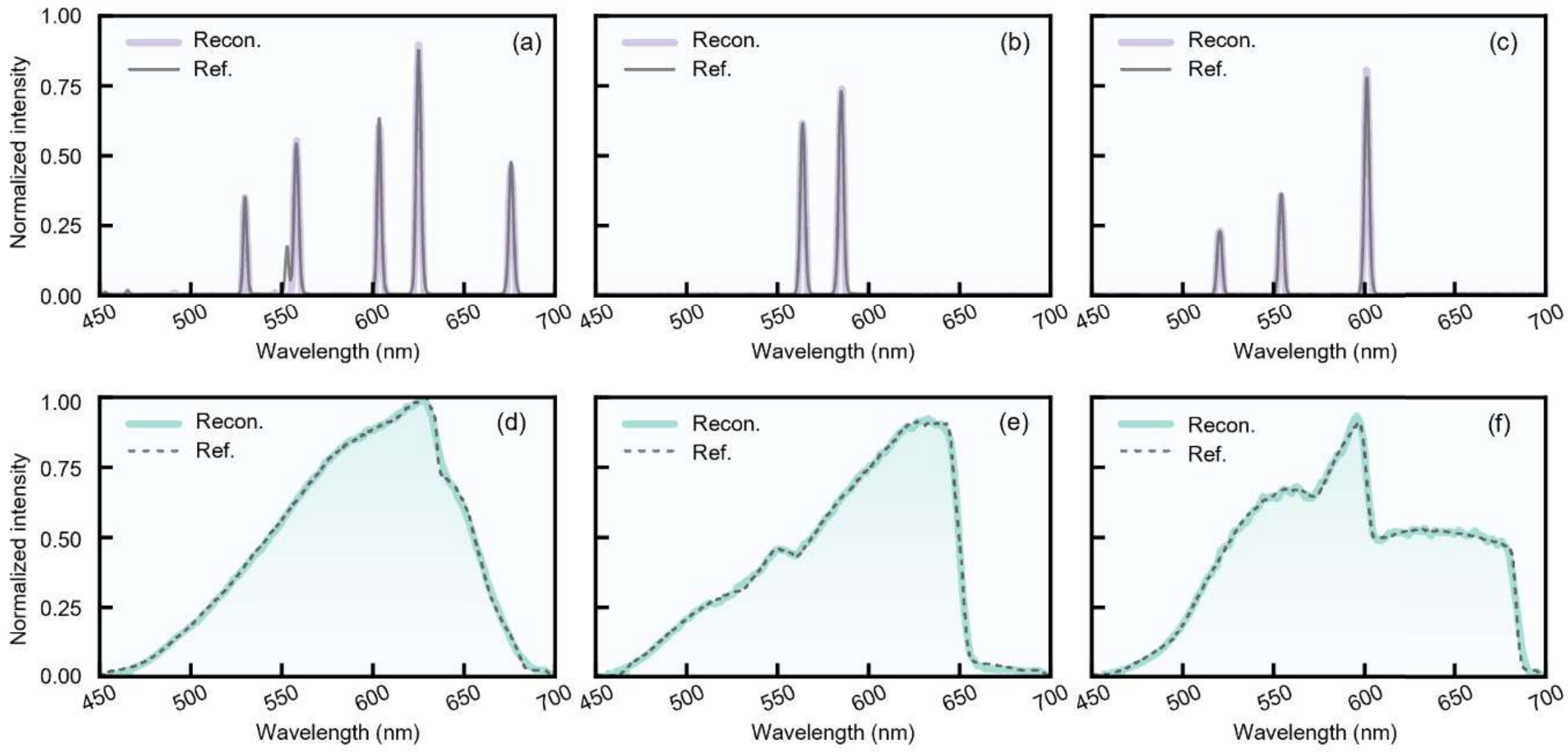

**Fig. S11. Additional recovery results from μSHADES for diverse input profiles.** The plots compare the ground truth spectra (Ref.) measured by the reference spectrometer with the spectra reconstructed by μSHADES (Recon.). (a–c) Reconstruction results for sparse, multi-peak signals. (d–f) Reconstruction results for continuous broadband spectral distributions. The system accurately captures the spectral envelope and intensity variations across the visible band, validating the robustness of the computational model against complex, broadband inputs

The system's performance on narrowband inputs is illustrated in Fig. S11a–c. For these discrete spectral peaks, the reconstruction algorithm demonstrates high localization accuracy, successfully resolving multiple closely spaced features across the 450–700 nm operating range. In addition, the recovered linewidths closely match the ground truth profiles without exhibiting significant spectral broadening or spurious artifacts.

In parallel, the system exhibits consistent performance for continuous broadband spectra, as shown in Fig. S11d–f. Unlike the sparse inputs, these profiles require the accurate recovery of low-frequency components and smooth transitions. The reconstruction tracks the complex modulation of the spectral envelope with high fidelity, showing close agreement between the ground truth and the predicted intensity distributions. The model effectively captures the overall shape and fine intensity variations without introducing baseline drift or amplitude bias, confirming its suitability for general-purpose spectroscopic applications.

## Supplementary Note 13. Test with a normal grating

To validate the efficacy of the proposed SGA design, we fabricated a control device based on the same optical design, with the only difference being the use of a normal dispersive grating. The control grating shares the same geometric parameters as μSHADES, including period and duty cycle, and differs only in that no stochastic initial phase is applied, which means the phase profile is strictly periodic.

The same experimental and calibration procedure is implemented for this normal grating spectrometer. The device is assembled and calibrated using the same setup as μSHADES, as detailed in Supplementary Note 11. The ASG is programmed to generate the same target spectra as those used in μSHADES to ensure a fair comparison. The resulting dataset again comprises 50,000 spectra, with 90% used for training and 10% reserved for validation. The same DL architecture as for μSHADES is applied to the normal grating spectrometer dataset, and all training hyperparameters, including learning rate, optimizer, and batch size, are kept identical.

A representative test case, showing the reconstruction of the same input spectrum from μSHADES and from the normal grating spectrometer, is presented in Fig. S12a, b. As evident from the comparison, the normal grating spectrometer exhibits markedly limited reconstruction fidelity relative to μSHADES. The peak positions and amplitudes recovered from the normal grating device display pronounced uncertainty and deviations. The corresponding pixel-wise absolute error curves in Fig. S12c further show that the reconstruction error of the normal grating spectrometer is substantially larger than that of μSHADES. Taken together, these results provide strong evidence for the superiority of the SGA architecture and substantiate our design choice of employing SGA in μSHADES.

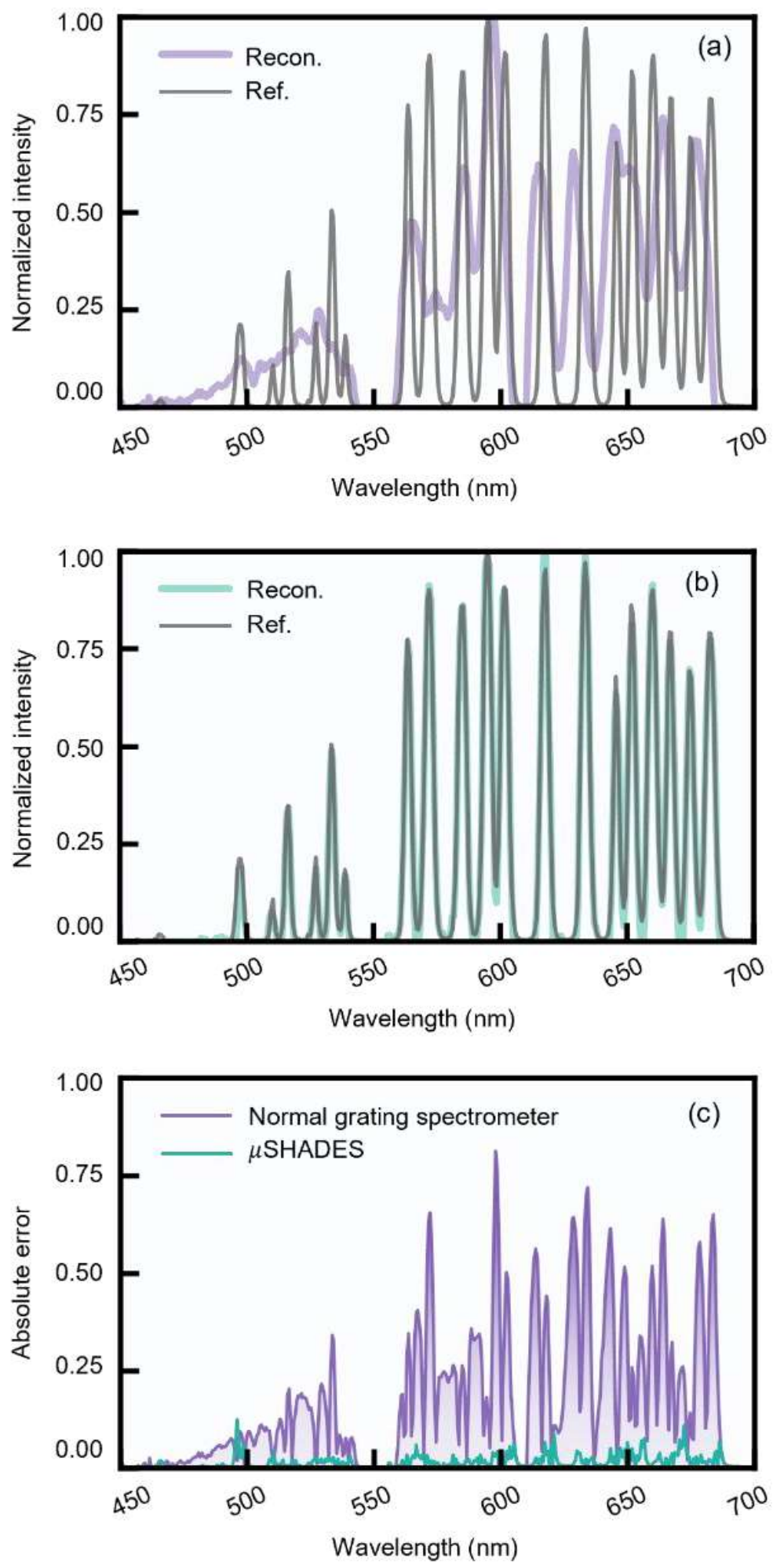

**Fig. S12. Comparison of spectrum reconstruction between the normal grating spectrometer and μSHADES.** (a) Reconstructed spectrum obtained with the normal grating spectrometer (Recon.) and the corresponding ground truth spectrum (Ref.). (b) Reconstructed spectrum (Recon.) obtained with the μSHADES for the same input, together with the ground truth spectrum (Ref.). (c) Pixel-wise absolute error for the two devices, showing that the μSHADES exhibits substantially smaller reconstruction error than the normal grating spectrometer.

## Supplementary Note 14. Laser-diffuser testing setup

To validate the impact of source coherence on the response of μSHADES, we establish a laser-diffuser testing platform, as illustrated in Fig. S13. A monochromatic laser source ($\lambda$ = 540 nm) is directed through a ground glass diffuser mounted on a motorized rotational stage. The transmitted light is subsequently collected by a coupling lens and focused into a liquid light guide, which delivers the illumination to the μSHADES device.

This setup allows for the controlled modulation of spatial coherence. As shown in Fig. S13a, when the rotational stage is held static, the laser light retains its high spatial coherence after passing through the stationary diffuser, generating a fixed, high-contrast speckle pattern on the sensor. This configuration serves as the testing baseline for a coherent light source. The experimental setup is detailed in Fig. S13b and Fig. S13c.

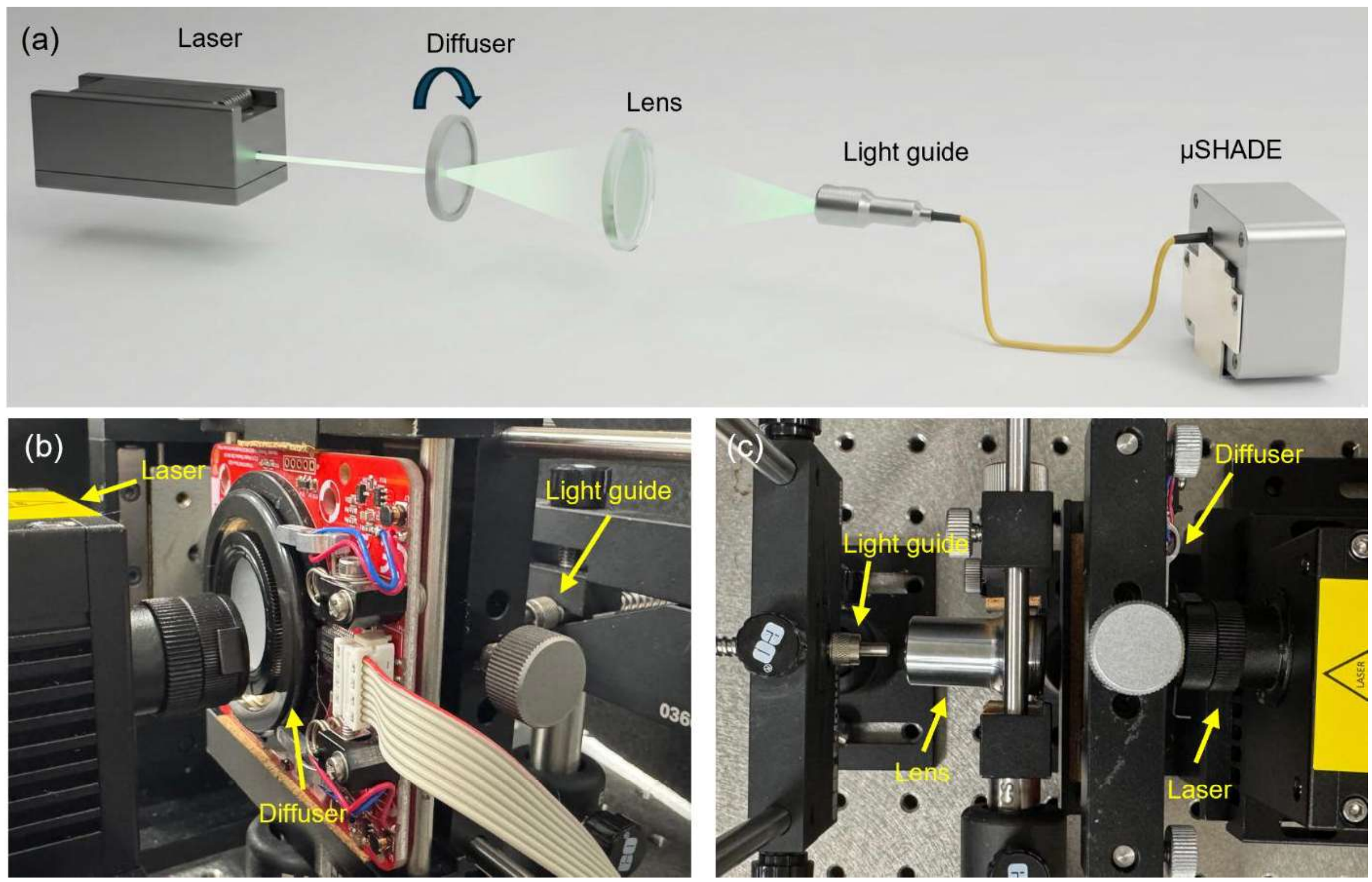

**Fig. S13. Laser-diffuser testing setup for coherence characterization. (**a) Schematic illustration of the optical path. A monochromatic laser beam is transmitted through a ground-glass diffuser placed before a coupling lens and the input of a light guide. The diffuser is either held static to preserve high spatial coherence or rotated to reduce the effective spatial coherence through temporal averaging during the camera exposure. The resulting spatially incoherent light is coupled into a light guide using a lens and delivered to the light guide, then to the μSHADES device for testing. (b, c) Photographs of the experimental implementation from different viewing angles, showing the alignment of the laser, the motorized rotational stage holding the diffuser, the collection lens, and the input light guide.

Conversely, to emulate an incoherent source that is used in ASG, the rotational stage is activated. As the laser spot traverses the diffuser rotating at a high tangential velocity, the phase delays induced by the scattering surface vary rapidly over time. Within the integration time of the sensor, these variations undergo temporal averaging, effectively removing the spatial coherence and smoothing out the speckle structure. As demonstrated in the main text, this removal of coherence allows the μSHADES system to correctly recover the spectral peak, verifying that the reconstruction algorithm remains accurate under incoherent illumination conditions.

## Supplementary Note 15. Evaluation of transfer-learned μSHADES

As demonstrated in the main text, transfer learning (TL) enables μSHADES to achieve reconstruction performance comparable to that of a model trained on the full calibration dataset while relying on a substantially smaller number of calibration measurements. Here we quantitatively evaluate this transfer-learned configuration. Specifically, only 10% of the paired calibration measurements are used for fine-tuning, mimicking a practical scenario in which limited training data are available for a new device or deployment. Representative reconstruction examples obtained from the transfer-learned model are shown in Fig. S14.

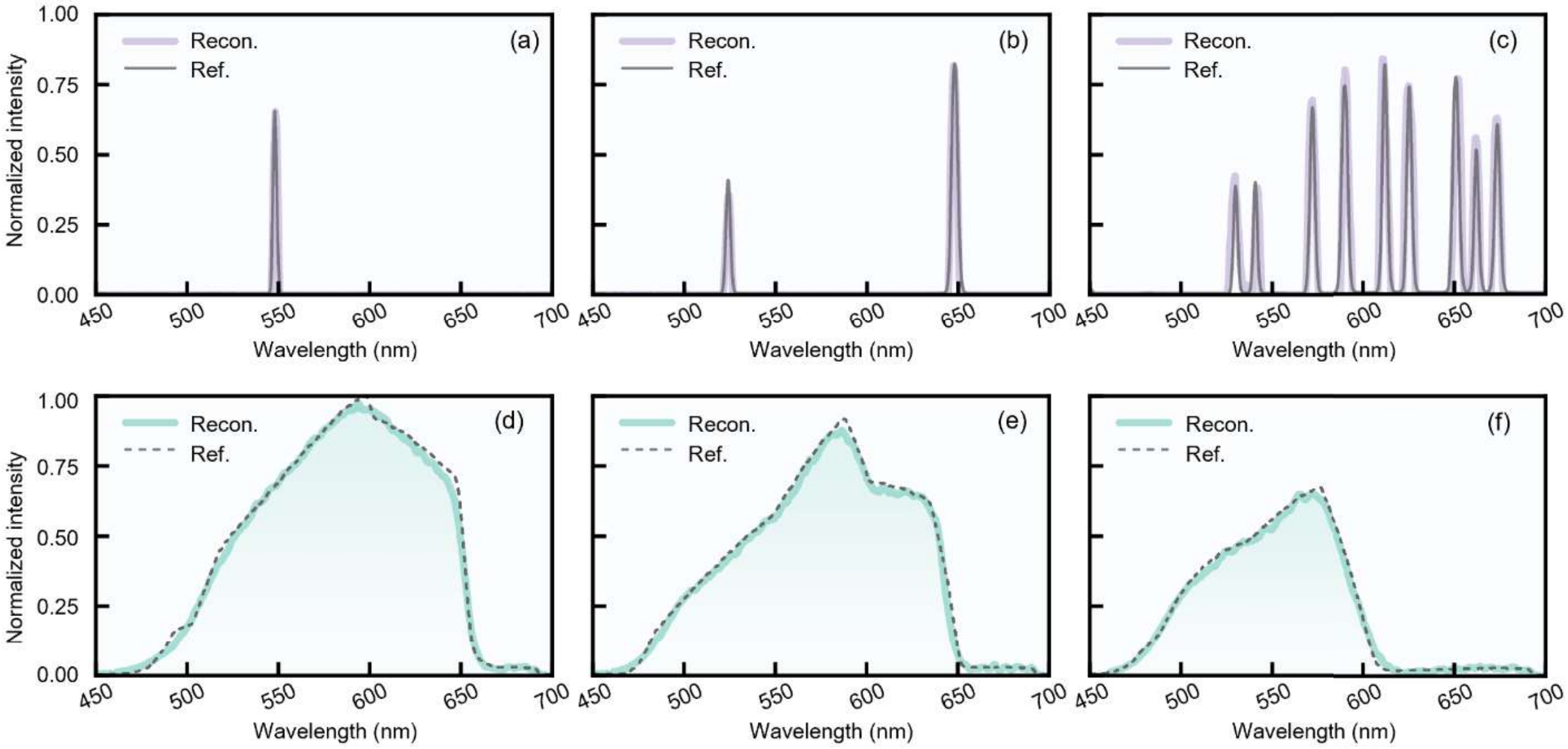

**Fig. S14. Representative reconstruction examples for the transfer-learned μSHADES.** The plots compare the ground truth spectra (Ref.) measured by the reference spectrometer with the spectra reconstructed by transfer-learned μSHADES (Recon.). (a–c) Reconstructed narrow-line spectra from the transfer-learned model compared with the corresponding ground truth spectra, showing single-peak, double-peak, and multi-peak cases, respectively. (d–f) Reconstructed broadband spectra with different envelopes from the transfer-learned model and the corresponding ground truth spectra, illustrating that high spectral resolution is largely preserved even when only 10% of the calibration dataset is used for TL.

During training, the loss for the TL model converges to a value close to that of the model trained on the full dataset, indicating that the network can still capture the essential spectral encoding of μSHADES under data-constrained conditions. Consistent with this behavior, the reconstructed spectra exhibit sharp, well-localized peaks. Taken together, these results confirm that TL enables μSHADES to maintain essentially the same reconstruction quality and spectral resolving power while substantially reducing calibration requirements, which is particularly advantageous when scaling to new devices and deployments.

## Supplementary Note 16. Comparison to state-of-the-art spectrometers

A comprehensive comparison between μSHADES and other reported compact spectrometers is provided in Table S1. Existing spectrometers often rely on dispersive or modulational elements such as gratings [3], photonic crystals [4], metasurfaces [5], nanowires [6], or stratified waveguides [7]. While these approaches can achieve competitive spectral resolutions, they are subject to intrinsic trade-offs, particularly in terms of optical throughput and fabrication complexity. As detailed in the table, many high-resolution designs exhibit low throughput due to limited input apertures or coupling inefficiencies. Also, their reliance on sophisticated nanofabrication techniques often leads to high manufacturing costs.

In contrast, μSHADES attains high spectral resolution while being inherently tolerant to optical aberrations, thereby alleviating the need for carefully corrected imaging optics and reducing system-level design complexity. The SGA is fabricated using a scalable lift-off process, which further simplifies production and lowers cost. At the same time, μSHADES achieves high optical throughput, addressing the throughput limitations common in slit-based or single-mode waveguide spectrometers. By combining this high-efficiency optical front-end with a deep-learning-based reconstruction framework, μSHADES delivers high spectral resolution suitable for low-light applications within a compact form factor.

The cost breakdown for a single μSHADES unit is summarized in Table S2. The dominant cost contributions arise from the reflective lens (approximately US$44) and the CMOS camera module (approximately US$30), both of which are mass-produced, off-the-shelf components. The random-grating wafer is self-fabricated and, at the current prototyping scale, is estimated to cost on the order of US$10 per device, while the mechanical housings contribute an additional approximately US$10. Consequently, the total cost per μSHADES device is approximately US$94. This cost is expected to decrease further in volume manufacturing, for example, through wafer-level processing of the SGA and batch production of the mechanical fixtures.

Beyond its low cost, μSHADES is highly competitive relative to existing off-the-shelf portable and miniaturized spectrometers, as detailed in Table S3. Compared with current compact instruments, μSHADES uniquely combines high spectral resolution, high throughput, a very small form factor, and low cost. For instance, commercial miniaturized spectrometers such as the Hamamatsu C12880MA exhibit limited optical throughput and modest spectral resolution, which restrict their applicability in more demanding use cases. Few-nanometer-resolution devices such as the Ocean Insight STS-VIS and the Avantes AvaSpec-Mini-series typically rely on conventional dispersive benches or miniaturized Czerny–Turner architectures, resulting in comparatively large device volumes and retail prices in the ~US$10,000 range. Handheld NIR spectrometers such as the VIAVI MicroNIR OnSite-W further integrate illumination and on-board electronics but remain substantially larger (>400 $cm^3$) and high-cost (US$20,000+). Within this landscape, μSHADES occupies a distinct design space by offering high-resolution, high-throughput performance in a compact and cost-efficient implementation.

In contrast, μSHADES offers a highly miniaturized 20 × 20 × 12 mm footprint while maintaining a spectral resolution of 2.4 nm, comparable to state-of-the-art commercial handheld high-resolution spectrometers. Importantly, all components in μSHADES are inexpensive and most are off-the-shelf, yielding a cost below US$100. This corresponds to a large reduction in cost and volume relative to existing instruments with similar resolutions. Moreover, the slit-free, SGA architecture of μSHADES provides a several-fold increase in optical throughput compared with conventional slit-based spectrometers, substantially expanding the range of viable application scenarios, particularly in low-light settings. The combination of performance, manufacturability, and cost positions μSHADES as a practical platform for large-scale deployment and various sensing applications.

**Table S1. Comparison between μSHADES and other compact spectrometers in the literature.**

| Design | Bandwidth (nm) | Resolution (nm) | Disperser | Reconstruction method | Throughput | Cost |
|---|---|---|---|---|---|---|
| Nat. Photonics (2013)[8] | 25 | 0.75 | Random medium | Optimization | Low | High |
| Nature (2015)[9] | 300 | 3 | Quantum dots | Optimization | Low | High |
| Optica (2016)[10] | 0.4 | 0.01 | Waveguide | Optimization | Low | High |
| Nat. Commun. (2018)[5] | 100 | 1.2 | Metasurface | Direct mapping | Medium | High |
| Nat. Commun. (2019)[4] | 200 | 1.0 | Photonic crystal slab | Optimization | Low | High |
| Science (2019)[6] | 130 | 7 | Nanowire | Optimization | Low | High |
| Nat. Photonics (2020)[11] | 500 | 5.5 | Thin film | Optimization | Medium | High |
| Nat. Commun. (2021)[7] | 180 | 0.45 | Waveguides | Optimization | Low | Low |
| Science (2022)[12] | 440 | 3 | 2D material | Optimization | Low | High |
| ACS Photonics (2022)[13] | 350 | 3 | MEMS | Optimization | Low | Medium |
| Adv. Sci. (2023)[3] | 400 | 5.8 | Grating | Direct mapping | Medium | Medium |
| eLight (2023)[14] | 550 | 0.00689 | Micro-taper | Direct mapping | Low | Medium |
| Nat. Commun. (2024)[15] | 330 | 0.65 | Microfilter array | Optimization | Medium | High |
| Light Sci. Appl. (2024)[16] | 200 | 3 | Pinhole | Optimization | Low | Low |
| Light Adv. Manuf. (2025)[17] | 330 | 5 | Planar lens | Optimization | Low | High |
| Nature (2024)[18] | 500 | 2.7 | Thin film | Deep learning | Low | High |
| **μSHADES (This work)** | **250** | **2.4** | **SGA** | **Deep learning** | **High** | **Low** |

**Table S2. Cost breakdown of µSHADES.**

| **Component** | **Cost/USD** | **Cat.no.** |
|---|---|---|
| CMOS module | $30 | Sony IMX586 |
| SGA and multi-slit wafer | $10 | Self-manufactured |
| Concave mirror | $44 | Edmund Optics #43-674 |
| Mechanical housing | $10 | Self-manufactured |
| **Total** | **$94** | |

**Table S3. Comparison with other commercial portable/miniaturized visible/Near-IR spectrometers.**

| Model | Vendor | Spectral range (nm) | Resolution (FWHM) | Footprint / Size (mm) | Estimated Volume ($cm^3$) |
|---|---|---|---|---|---|
| C12880MA | Hamamatsu | 340–850 | ~12 nm (typical), 15 nm (maximum) | 20.1 × 12.5 × 10.1 | ~2.5 |
| C12666MA | Hamamatsu | 340–780 | ~12 nm (typical), 15 nm (maximum) | 20.1 × 12.5 × 10.1 | ~2.5 |
| C11708MA | Hamamatsu | 640–1050 | ~15 nm (typical), 20 nm (maximum) | 27.6 × 16.8 × 13 | ~6.0 |
| STS-VIS | Ocean Insight | 350–800 | ~1–12 nm (depends on slit) | 40 × 42 × 24 | ~41 |
| AvaSpec-Mini 2048CL | Avantes | 200–1100 | 0.10–22 nm (configuration dependent) | 95 × 68 × 20 | ~129 |
| DLP NIRscan Nano | Texas Instruments | 900–1700 | 10–12 nm | 62 × 58 × 36 | ~130 |
| AvaSpec-Mini-NIR256-1.7 | Avantes | 900–1750 | ~2 nm (configuration dependent) | 95 × 68 × 20 | ~129 |
| **µSHADES** | **N/A** | **450–700 (extendable)** | **2.4 nm** | **20 × 20 × 12** | **4.8** |

Notes: Specifications obtained from manufacturers' datasheets and distributor product pages (Hamamatsu, Ocean Insight, Avantes, Texas Instruments; accessed 2024–2025).